\documentclass[sigconf, nonacm]{acmart}
\AtBeginDocument{%
  \providecommand\BibTeX{{%
    \normalfont B\kern-0.5em{\scshape i\kern-0.25em b}\kern-0.8em\TeX}}}

\setcopyright{none}

\acmConference[ICSE 2024]{46th International Conference on Software Engineering}{April 2024}{Lisbon, Portugal}

\def\name{{TensorShield}}

\usepackage{subfigure}
\usepackage{tikz}
\usepackage{algpseudocode}
\usepackage{enumerate}

\usepackage{tabularray}
\usepackage{booktabs}
\usepackage{tabularx}
\usepackage{stfloats} 
\newcolumntype{Y}{>{\centering\arraybackslash}X} 
\usepackage{pifont} 
\usepackage{bbding} 
\usepackage{makecell} 
\usepackage{multirow} 
\usepackage{listings} 

\usepackage[linesnumbered,boxed,ruled,commentsnumbered,vlined]{algorithm2e} 

\newcommand{\STAB}[1]{\begin{tabular}{@{}c@{}}#1\end{tabular}} 

\newcommand{\tabincell}[2]{\begin{tabular}{@{}#1@{}}#2\end{tabular}} 
\usepackage{bm}		
\usepackage[T1]{fontenc}
\usepackage{enumitem} 
\setenumerate[1]{itemsep=0pt,partopsep=0pt,parsep=\parskip,topsep=2pt} 
\setitemize[1]{itemsep=2pt,partopsep=0pt,parsep=\parskip,topsep=2pt} 
\setdescription{itemsep=0pt,partopsep=0pt,parsep=\parskip,topsep=2pt} 

\usepackage{threeparttable} 
\usepackage{tikz} 
\usepackage{colortbl} 
\usepackage{arydshln} 
\usepackage{url}

\definecolor{codegreen}{rgb}{0,0.6,0}
\definecolor{codepurple}{rgb}{0.58,0,0.82}
\lstdefinelanguage{DLANG}{
  sensitive = true,
  tabsize=2,
  backgroundcolor=\color{white},
  basicstyle=\linespread{1}\ttfamily\footnotesize,
  keywords={FROM, FUNC, TZ_DATA_BACKWARD, TZ_FUNC, TZ_STORE,TZ_FUNC_SUB, TZ_DATA_FORWARD, TZ_DATA_ONLY, TZ_DATA_INTG, TZ_DATA_CONF},
  keywordstyle=\color{blue!90}\bfseries,
  otherkeywords={
    >, <, ==
  },
  keywords = [2]{int, for, char, if, else, while},
  keywordstyle=[2]\color{magenta},
  keywords = [3]{TZ_genKey, TZ_sign_rsa_sha1, TZ_enc_aes_cbc},
  keywordstyle=[3]\color{blue!90}\bfseries,
  keywords = [4]{@DRIVER, INSERT_AFTER, END_INSERT},
  keywordstyle=[4]\color{blue!90}\bfseries\itshape,
  keywords = [5]{STRING_TZ, INT_TZ},
  keywordstyle=[5]\color{blue!90}\bfseries,
  commentstyle=\color{codegreen},
  numbers=left,
  numberstyle=\footnotesize,
  stepnumber=1,
  numbersep=4pt,
  showstringspaces=false,
  breaklines=true,
  frame=lines,
  comment=[l]{//},
  morecomment=[s]{/*}{*/},
  postbreak=\mbox{\textcolor{red}{$\hookrightarrow$}\space},
}
\newcommand\bluecmt[1]{{\color{blue}{#1}}}
\def\banduibancuo{\ding{51}\rotatebox[origin=c]{-9.2}{\kern-0.7em\ding{55}}}

\usepackage{tikz}
 
\newcommand*\halfcirc[1][1ex]{%
	\begin{tikzpicture}
	\draw (0,0) circle (#1);
	\draw[fill] (0,0)-- (90:#1) arc (90:270:#1) -- cycle ;
	\end{tikzpicture}}
\newcommand*\fullcirc[1][1ex]{\tikz\fill (0,0) circle (#1);}

\newcommand*\colourcheck[1]{%
  \expandafter\newcommand\csname #1check\endcsname{\textcolor{#1}{\ding{52}}}%
}
\newcommand*\colourcross[1]{%
  \expandafter\newcommand\csname #1cross\endcsname{\textcolor{#1}{\ding{55}}}%
}
\colourcheck{teal}

\colourcross{blue}
\colourcross{green}
\colourcross{mypink}
\colourcross{pink}
\colourcross{salmonpink}
\definecolor{mypink}{RGB}{251,229,214}
\definecolor{myblue}{RGB}{218,227,243}
\definecolor{salmonpink}{rgb}{1.0, 0.57, 0.64}
\definecolor{mygrey}{RGB}{239,239,239}
\definecolor{Gray}{RGB}{239,239,239}
\newcolumntype{a}{>{\columncolor{Gray}}c}
\newcolumntype{b}{>{\columncolor{white}}c}

\begin{document}

\title{\name: Safeguarding On-Device Inference by Shielding Critical DNN Tensors with TEE}


\author{Tong Sun$^1$, Bowen Jiang$^1$, Hailong Lin$^1$, Borui Li$^2$, Yixiao Teng$^1$, Yi Gao$^1$, and Wei Dong$^1$}
\affiliation{
  \institution{$^1$The State Key Laboratory of Blockchain and Data Security\\
College of Computer Science, Zhejiang University, China\\
$^2$School of Computer Science and Engineering, Southeast University, China}
  \streetaddress{}
  \city{}
  \state{}
  \country{}
  \postcode{}
}
\email{{tongsun,jiangbw,linhl,tengyixiao, gaoyi,dongw}@zju.edu.cn,libr@seu.edu.cn}
\settopmatter{printfolios=true}
\settopmatter{printacmref=false}


\begin{abstract}
To safeguard user data privacy, on-device inference has emerged as a prominent paradigm on mobile and Internet of Things (IoT) devices.
This paradigm involves deploying a model provided by a third party on local devices to perform inference tasks. 
However, it exposes the private model to two primary security threats: model stealing (MS) and membership inference attacks (MIA). 
To mitigate these risks, existing wisdom deploys models within Trusted Execution Environments (TEEs), which is a secure isolated execution space. 
Nonetheless, the constrained secure memory capacity in TEEs makes it challenging to achieve full model security with low inference latency.

This paper fills the gap with \textbf{TensorShield}, the first efficient on-device inference work that shields partial tensors of the model while still fully defending against MS and MIA.
The key enabling techniques in TensorShield include:
(i) a novel eXplainable AI (XAI) technique exploits the model's attention transition to assess critical tensors and shields them in TEE to achieve secure inference, and
(ii) two meticulous designs with critical feature identification and latency-aware placement to accelerate inference while maintaining security.
Extensive evaluations show that TensorShield delivers almost the same security protection as shielding the entire model inside TEE, while being up to 25.35$\times$ (avg. 5.85$\times$) faster than the state-of-the-art work, without accuracy loss.

\end{abstract}

\maketitle

\section{Introduction}\label{sec:introduction}
\begin{figure}
    \centering
    \includegraphics[width=\linewidth]{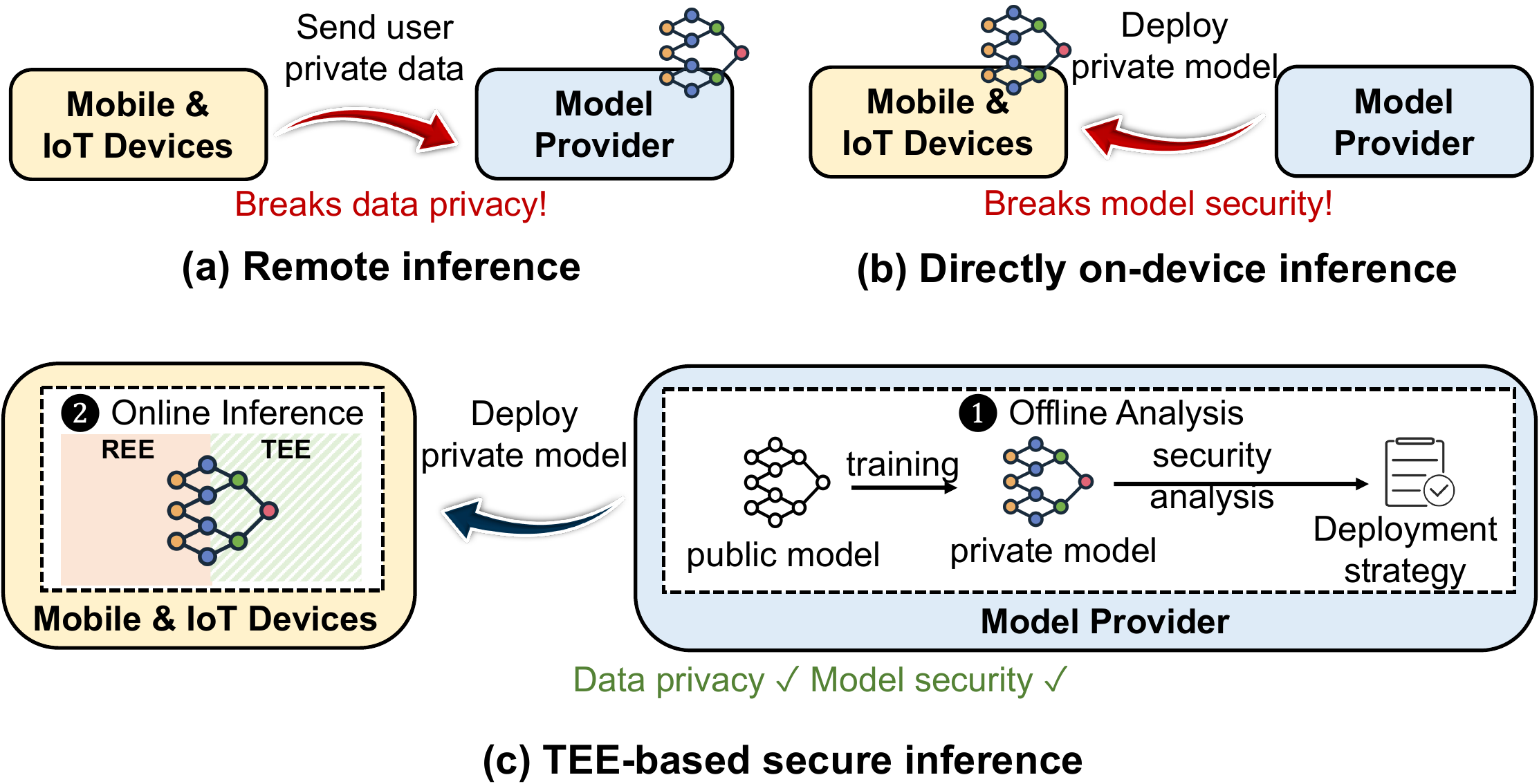}
    \vspace{-2em}
    \caption{Paradigms of interactions between user and private models. (a) Remote inference: users send data to the model provider. (b) Directly on-device inference: model providers deploy the private model in devices. (c) TEE-based secure on-device inference: model providers generate a secure deployment strategy (\ding{182}) and deploy the model with TEE (\ding{183}).  }
    \label{fig:TEEsolutions}
    \vspace{-2em}
\end{figure}

On-device inference has become an important paradigm for privacy-  and latency-sensitive tasks on mobile and Internet of Things (IoT) devices~\cite{yuan2022infi,fang2018nestdnn,wang2021asymo,guo2021mistify,chien2023enc2, jia2022codl, zhang2023comprehensive}.
Recent surveys~\cite{sun2021mind,xu2019first} have indicated that numerous mobile applications have implemented on-device deep learning models (DNNs) for a variety of uses, including liveness detection, video processing, and face recognition.
The major advantages of on-device inference are obvious: (i) it protects user data privacy. As shown in Figure~\ref{fig:TEEsolutions}(a), remote inference requires transmitting user sensitive data to the untrusted clouds, compromising data privacy, (ii) it reduces the delays inherent in network communications, and (iii) it operates independently of internet connectivity. 

However, on-device inference introduces new security threats to the deployed private deep neural network (DNN) models on billions of devices.
As shown in Figure~\ref{fig:TEEsolutions}(b), attackers can easily obtain model weights and inference information (e.g., by dumping device memory), facilitating representative attacks, such as model stealing (MS) and membership inference attacks (MIA) at low costs.

In MS attacks, the adversary queries the target model with carefully crafted samples to maximally extract internal model information and then uses the returned query results to train a surrogate model~\cite{shen2022model, orekondy2019knockoff}.
MS enables attackers to obtain a surrogate (a.k.a., shadow) model that closely replicates the functionality of the deployed model, leading to the leakage of private assets of the model provider. 
MIA allows attackers to query whether a sample is in the training set, resulting in sensitive data leakage~\cite{chen2022relaxloss, mo2020darknetz, jia2019memguard}.
These attacks could lead to severe security threats to numerous IoT applications.
For example, attackers can launch MS attacks to obtain a surrogate model so as to construct adversarial samples, misleading the vision recognition systems of autonomous vehicles.
Attackers can also launch MIA attacks to a model on health monitoring devices, causing privacy leakage of the user’s health conditions. 

Recently, TEE-based secure inference approaches attract much research attention~\cite{lee2019occlumency, zhanggroupcover, sun2023shadownet, shen2022soter, elgamal2020serdab, mo2020darknetz, zhang2023no, sun2024tsqp, hou2021model}, primarily due to its high efficiency compared with previous security approaches such as Homomorphic Encryption (HE)~\cite{kim2023optimized, xiao2021privacy, chen2019efficient}. 
In TEE-based secure inference approaches (Figure~\ref{fig:TEEsolutions}(c)), the model provider can train their private model from a large public model using the private dataset. 
The training process can be monitored for security analysis. Given the private model, different security approaches may then conduct different strategies to safeguard the inference process. 

\begin{figure*}[t]
    \centering
    \includegraphics[width=\linewidth]{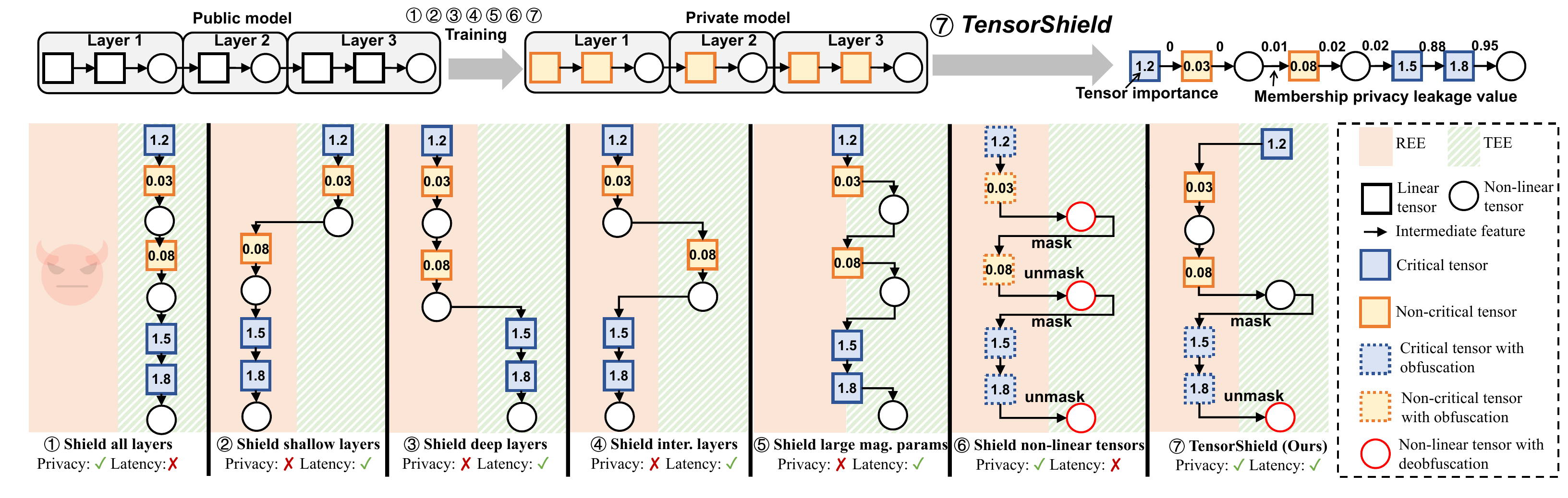}
    \vspace{-2.5em}
    \caption{An illustration of previous work for model protection. The model has three layers with eight tensors. \ding{172} shields all (three) layers~\cite{lee2019occlumency}. \ding{173} shields one shallow layer (Layer1)~\cite{elgamal2020serdab} and \ding{174} shields one deep layer (Layer3)~\cite{mo2020darknetz}. \ding{175} shields one random intermediate layers (Layer2)~\cite{shen2022soter}. \ding{176} shields large magnitude weights of each layer~\cite{hou2021model}. \ding{177} shields non-linear layers and obfuscates all linear tensors (indicated by rectangles with dashed lines) and all intermediate features~\cite{sun2023shadownet, zhanggroupcover}. \name~(Ours) shields critical tensors (indicated by blue rectangles) and only masks privacy-related intermediate features.}
    \label{fig:ExistingWork}
    \vspace{-1em}
\end{figure*}
Some approaches achieve secure inference by shielding all layers. For example, Occlumency~\cite{lee2019occlumency} (\ding{172} of Figure~\ref{fig:ExistingWork}) shields all layers by executing the entire model in the TEE, using techniques such as on-demand loading of the weights and partitioned convolution calculation to address TEE’s memory limitation. 
ShadowNet~\cite{sun2023shadownet} and GroupCover~\cite{zhanggroupcover} (\ding{177} of Figure~\ref{fig:ExistingWork}) also shield all layers by placing all non-linear layers in the TEE. The linear layers can be protected in REE with obfuscating by matrix transformation. These approaches can cause a high execution overhead because of on-demanding or (de)obfuscation. 

To reduce the execution overhead of on-device inference, some other approaches only shield partial layers. 
For example, Serdab~\cite{elgamal2020serdab} (\ding{173} of Figure~\ref{fig:ExistingWork}) shields the shallow layers in the TEE to protect general information.  DarkneTZ~\cite{mo2020darknetz} (\ding{174} of Figure~\ref{fig:ExistingWork}) shields the deep layers in the TEE to protect classifier information. Unfortunately, these approaches cannot fully defend against MS and MIA attacks. The underlying reason lies in that they often fail to shield the most critical part with respect to model security. 
The essential of MS is to steal the \textit{decision capabilities} of the model.
For example, in DNN for image classification, the decision capability determines the category of an image.
If these capabilities are well shielded, we can fully defend against MS and MIA \cite{zhang2023no}.
However, the decision capabilities of a model do not necessarily depend on the shallow layers or deep layers. 
The \textit{key idea} of TensorShield is to identify the most critical part of a model at the granularity of tensors, which offers a finer granularity than layers. 
By shielding these critical tensors (in TEE or in REE via obfuscation), we can achieve a high level of security without a large execution overhead. 

How to identify the critical tensors? We rely on the recent progress in eXplainable AI (XAI) technique~\cite{huang2022real,huang2023elastictrainer} which leverages the contribution of weight and gradient changes to evaluate tensor importance.
However, directly employing XAI incurs inaccuracy for model security evaluation because it only focuses on the contribution to the loss function and ignores the evaluation of decision-making ability (cf. \S\ref{subsec:design_critical_tensor}). 
To this end, we propose a new metric called attention transition between the public pre-trained model and the victim model to accurately evaluate tensor criticality (cf. \S\ref{subsec:design_critical_tensor}). 
After assessing its criticality, \name\ trains a surrogate model on the validation dataset to simulate an attacker. 
We leverage the convergence speed of the surrogate model to select critical tensors in the TEE (cf. \S\ref{subsec:design_critical_tensor_selector}).

We have further proposed two novel techniques to reduce the inference latency.
(i) instead of previous work that masks all transmitted features between the TEE and REE via one-time padding (OTP), TensorShield only masks privacy-related features determined by the JS-divergence distance to reduce latency overhead of masking (cf. \S\ref{subsec:design_selective_masking}), and (ii) unlike previous efforts that primarily leverage the REE GPU for inference, TensorShield conducts fine-grained modeling of hardware platform capabilities to minimize inference latency (cf. \S\ref{subsec:design_modeling}).

We evaluate the TensorShield with four well-deployed
DNN models and four datasets on two hardware platforms (cf. \S\ref{sec:evaluation}).
Compared to existing approaches that shield partial layers~\cite{mo2020darknetz, elgamal2020serdab}, TensorShield reduces MS and MIA accuracy by up to 45.97\% (avg. 28.54\%) and 38.33\% (avg. 23.15\%), respectively, and reduces inference latency by up to 7.17$\times$ (avg. 1.49$\times$). 
In comparison to approaches that shield all layers~\cite{lee2019occlumency, zhanggroupcover}, TensorShield can reduce inference latency by up to 25.35$\times$ (avg. 5.85$\times$) and achieves almost the same level of protection.

We summarize our contributions as follows:
\begin{itemize}[leftmargin=*]
    \item We propose a novel XAI technique for critical tensor identification to defend against MS. We propose a metric named attention transition to more accurately evaluate the criticality of tensors. 
    \item We propose a critical feature identification to defend against MIA. We leverage JS-divergence to assess membership privacy leakage and selective mask features via the OTP.
    \item We propose a latency-aware placement technique, the first to determine the placement of shielded tensors by jointly considering the hardware capabilities and the criticality of tensors and features.
    \item Our thorough evaluation shows that TensorShield offers almost the same security guarantee as shielding the entire DNN models inside TEE with over 25$\times$ inference performance improvements than the state-of-the-art works.
\end{itemize}

\begin{figure}[t]
    \centering
    \includegraphics[width=\linewidth]{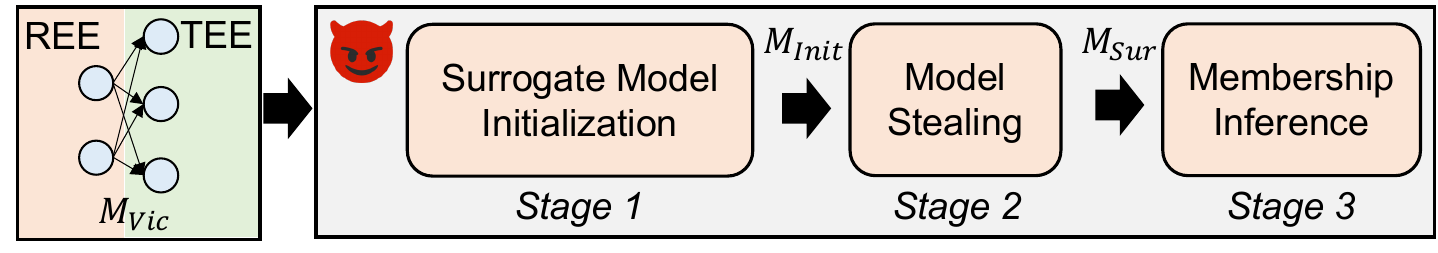}
    \vspace{-2.5em}
    \caption{A three-stage attack pipeline.}
     \vspace{-2em}
    \label{fig:attack_pipeline}
\end{figure}
\section{Background and Threat Model}
\subsection{Background}\label{sec:background}
\textbf{Trusted Execution Environment.} 
TEE (e.g., ARM TrustZone~\cite{trustzone}) provides a physical isolation scheme in the hardware devices that separates memory into the REE and TEE, where the REE can communicate with the TEE by invoking a secure monitor call.
This setup ensures that only legitimate users can access the secure world, while attackers are blocked.


\textbf{On-device inference attacks.}
On-device inference can fully preserve data in-situ, making it a widely utilized approach in privacy-sensitive applications~\cite{yuan2022infi,fang2018nestdnn,wang2021asymo,guo2021mistify,chien2023enc2, jia2022codl, zhang2023comprehensive}.
However, deploying third-party private models on devices poses security risks~\cite{liu2022ml, mo2024machine}.
As shown in Figure~\ref{fig:attack_pipeline},
two prominent types of inference attacks that threaten model privacy are model stealing (MS) and membership inference attacks (MIA). 
First, the attacker infers the architecture of the victim model based on the REE part and the model output with existing
techniques.
Then, the attacker chooses a public
model $\mathcal{M}_{Pub}$ (with the same architecture)
as $\mathcal{M}_{Init}$.
Lastly, the attacker transports the model weights in the unshielded (i.e., REE) part of $\mathcal{M}_{Vic}$ to the corresponding parts of
$\mathcal{M}_{Init}$.

The goal of model stealing (a.k.a., model extraction) is to extract the parameters from a target (a.k.a., victim) model.
The adversary queries the target model with carefully crafted samples to maximally extract internal model information and then uses the returned query results to generate an auxiliary dataset and train a surrogate model.
Ideally, an adversary will be able to obtain a surrogate (a.k.a., shadow) model with very similar performance as the target model. 
More formally:
\begin{equation}
    \texttt{MS: }\mathcal{M}_{Vic},\mathcal{D}_{aux} \rightarrow \mathcal{M}_{Sur}  \nonumber
\end{equation}
where $\mathcal{M}_{Vic}$ is the victim model, $\mathcal{M}_{Sur}$ is the surrogate model, and $\mathcal{D}_{aux}$ is an auxiliary dataset.

MIAs pose serious privacy risks, leading to extensive research. 
These attacks aim to identify whether a particular sample was used in training a model. 
In black-box MIAs, attackers use model outputs and auxiliary data without accessing internal details. 
In contrast, white-box MIAs exploit both model outputs and internal information like gradients and activations to improve their effectiveness.
More formally, given a target data sample $x_{target}$, and a stolen model $\mathcal{M}_{Sur}$, and an auxiliary dataset $\mathcal{D}_{aux}$, an MIA can be defined as: 
\begin{equation}
    \texttt{MIA: }x_{target}, \mathcal{M}_{Sur},\mathcal{D}_{aux} \rightarrow \{\textit{\text{member}}, \textit{\text{non-member}}\}.  \nonumber
\end{equation}
The term MIA accuracy refers to the accuracy of the surrogate model $\mathcal{M}_{Sur}$ for the member prediction task.
The term MS accuracy refers to the accuracy of the surrogate model $\mathcal{M}_{Sur}$ for the victim's original task.
Thus, the lower these two accuracies, the more effective the protection is considered. 
In this paper, we use $Acc_{MS}^{AllShield}$ and $Acc_{MIA}^{AllShield}$ to represent the MS accuracy and MIA accuracy achieved by shielding the entire model with TEE, which represents the lower-bound level of security protection.


\subsection{Threat Model}
\textbf{Device platform.}
In this paper, we focus on mobile and IoT devices requiring low-overhead protection approaches. 
We assume these devices are equipped with TEE, such as ARM TrustZone, which is prevalent in mobile and IoT devices~\cite{pinto2019demystifying}.
Furthermore, we assume the TEE ensures the protection of data privacy and code execution against unauthorized access. 
Conversely, the REE is completely under the control of attackers. 
Side-channel attacks~\cite{yuan2024ciphersteal, yuan2024hypertheft, bukasa2018trustzone, ryan2019hardware, 274707, deng2023cipherh} that could potentially lead to the leakage of sensitive information from the TEE are outside the scope of our study.

\textbf{Defender.} 
In our context, the defender is the model provider. 
Consistent with previous TEE-based secure inference methods, our goal to protect the deployed private model from unauthorized access and to prevent training dataset privacy breaches. 
Specifically, this paper aims to degrade white-box attacks to label-only black-box attacks, effectively achieving the protection level of enclosing the entire model within the TEE. 
We adopt the security assumption of a black-box baseline where the TEE fully shields the DNN model and returns only prediction labels, considered the upper-bound level of security protection offered by previous approaches.
Following recent literature~\cite{zhang2023no}, we do not seek to entirely prevent information leakage from TEE outputs, such as prediction labels. 
To ensure high-quality ML services, the model provider ensures the accuracy of models. 
The system detects any fault injections in parameters or modifications in inference results that could compromise integrity. 
Advanced protective techniques safeguard the model before it is securely transmitted to edge TEEs using robust encryption and authentication methods.
Once transmitted, the TEE selectively offloads the necessary model segments into the REE. 
We consider the transmission and offloading processes secure and tamper-proof~\cite{tramer2018slalom}. 
As the model is deployed on the edge, protecting users’ private inputs is unnecessary.

\textbf{Adversary.} 
We assume the attacker possesses considerable power, enabling control over all aspects of the device environment excluding the TEE. 
We recognize that effective security mechanisms should not depend on the secrecy of the method itself, hence, the attacker is fully informed about the protection strategy implemented by the defender. 
We assume that both the model owner and the attacker can use the public model $\mathcal{M}_{Pub}$ on the Internet to improve the accuracy of the model or attacks, a realistic setting for modern on-device learning tasks. 
Consistent with previous work, we assume the private (i.e., victim) model is trained based on a pre-trained public model.
The attacker can infer the architecture of the whole protected model, or an equivalent one, based on the public information (e.g., inference results or the unprotected model part). 
Besides, we assume that the attacker can query the victim model for limited times ($\leq$ 1\% of the training data), a practical assumption shared by previous work~\cite{zhang2023no, sun2024tsqp}. 

\section{System Overview}\label{sec:overview}
\begin{figure*}[t]
    \centering
    \includegraphics[width=.95\linewidth]{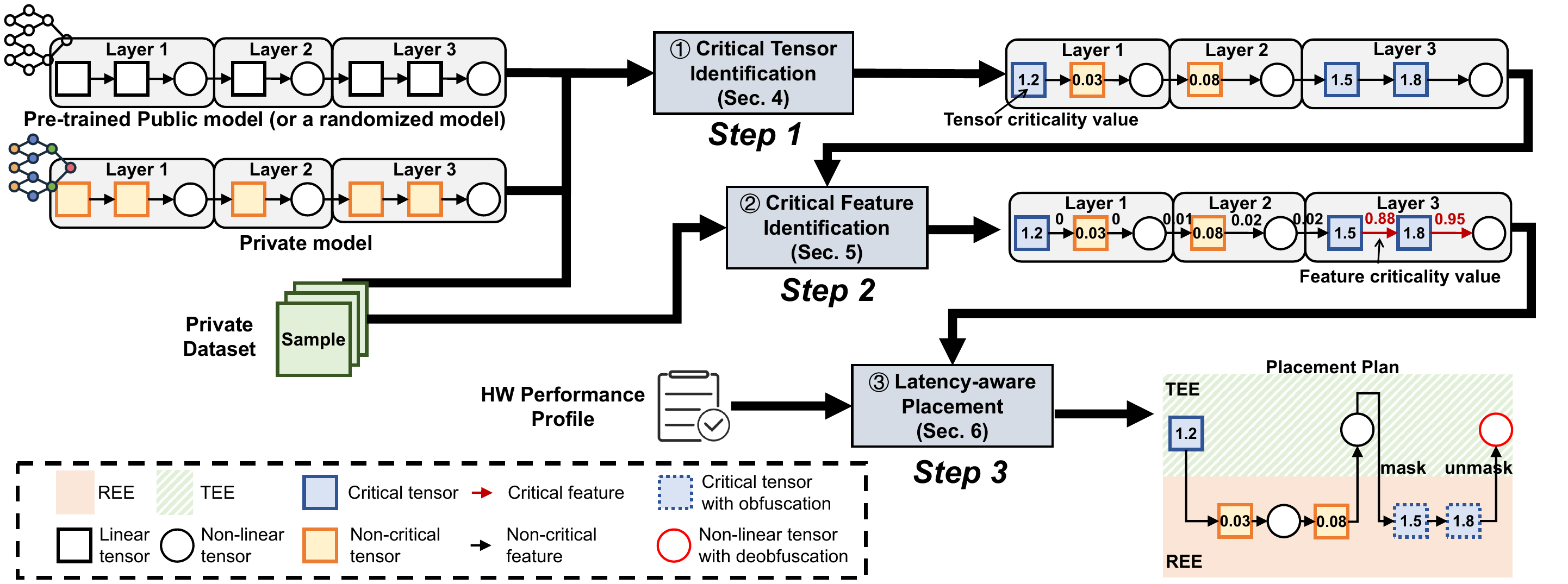}
    \vspace{-1em}
    \caption{Workflow of \name.}
    \label{fig:overview}
    \vspace{-1em}
\end{figure*}
\bluecmt{

}
Figure~\ref{fig:overview} shows the workflow of TensorShield. 
For protection, we first need to identify the critical tensors (cf. \S\ref{sec:critical_tensor_identification}) given the private model as well as the private dataset and the pre-trained public model. 
Each linear tensor (indicated by a rectangle) obtains a criticality value as shown in Figure~\ref{fig:overview}(\ding{172}). 
If a tensor has a large criticality value, it has a large contribution to the final model stealing accuracy. Note that non-linear tensors (indicated by circles) do not obtain criticality values. 
This is because they do not contain trainable parameters so they do not require protection. 
We then select a set of critical tensors (indicated by blue rectangles) for protection so that the model stealing accuracy does not exceed a given threshold (e.g., $Acc^{AllShield}_{MS}+1\%$). 

We next need to evaluate the criticality value of each cross-tensor transmission (i.e., the intermediate feature exchanged between two subsequent two tensors), as shown in Figure~\ref{fig:overview}(\ding{173}). 
We then select a set of critical intermediate features (cf. \S\ref{subsec:design_selective_masking}) so that the membership inference attack accuracy does not exceed a given threshold, e.g., $Acc^{AllShield}_{MIA}+1\%$. 
These intermediate features require masking and unmasking if the transmission across the TEE and REE boundary. 
No action will be taken for the intermediate features if the two subsequent tensors are placed at one side.

Finally, we need to decide how to place the tensors by jointly considering the criticality value of each tensor and intermediate features as well as the hardware performance profile (cf. \S\ref{subsec:design_modeling}), as shown in Figure~\ref{fig:overview}(\ding{174}). 
Each critical linear tensor can be placed in TEE or in REE via obfuscation, depending on the execution performance. 
The placement of the non-linear tensor depends on its previous linear tensor: if the previous linear tensor is placed in REE by obfuscation, the non-linear tensor should be placed in TEE because the protection via obfuscation only takes effect for linear tensors. 
If the transmission between two linear tensors is critical and the linear tensor is placed in REE, then its preceding and succeeding non-linear tensors should be placed in TEE for masking and unmasking. Transmission across the TEE and REE should be protected if it is identified as critical.

It is worth noting to emphasize some major differences compared to existing works:
(1)	Achieving the same level of protection, TensorShield protects far more less tensors. 
In our example, we only need to protect three linear tensors while shielding all layers approaches need to protect all eight tensors.
(2)	TensorShield selectively masks intermediate features related to membership privacy. 
For example, we only perform one pair of masking and unmasking operations while existing works require two pairs of masking and unmasking operations for all intermediate features.
(3)	The execution of critical tensors can be either in TEE or REE via obfuscation depending on the execution efficiency. 
For example, the first tensor is placed in TEE while the sixth and seventh tensors are placed in REE via obfuscation. 
Note that mobile GPU has limited computation capability, rendering execution in TEE without obfuscation more computationally efficient.

\section{XAI-based Critical Tensor Identification}\label{sec:critical_tensor_identification}
\subsection{Criticality Value Evaluation}\label{subsec:design_critical_tensor}
As we have described in \S\ref{sec:overview}, we have to identify a set of critical tensors to be protected, i.e., either in TEE or in REE via obfuscation. 
We say that a set of tensors to be critical if we can use the corresponding parameters and structures to train a surrogate model which leads to a high model stealing accuracy. 
Consider a victim model’s accuracy is 90\%. Protecting the entire model may lead to an accuracy of 60\%. A surrogate model may achieve a high accuracy of 85\%. 

Given a set of protected tensors, we can measure the model stealing attack accuracy by simulating the attack. The simulation involves the following steps: (1) Randomly selecting a small sample (less than 1\% of the training set) from the victim model’s validation dataset and erasing labels, then imputing it to the victim model to generate a pseudo-labeled dataset. (2) Using a pre-trained public model with the same structure as the victim model as the surrogate model, and initializing the unshielded victim model’s tensors to the corresponding surrogate model. (3) Training the surrogate model on the pseudo-labeled dataset (typically for 100 epochs). (4) Testing the surrogate model accuracy with the validation dataset as the model stealing accuracy.

However, identifying the set of critical tensors requires testing all possible combinations of tensors, which is always infeasible since well-deployed DNN models usually contain a large number of tensors. 

We would like to exploit the recent advances in eXplainable AI (XAI)~\cite{huang2022real, huang2023elastictrainer} to give each tensor a criticality value which correlates the corresponding tensor to the final model stealing accuracy. 
Intuitively, if a tensor has a large criticality value, it has a large contribution to the final model stealing accuracy. 

Existing efforts~\cite{zhang2022all, huang2023elastictrainer} suggest that tensor influences are not equally significant. 
Specifically, some tensors are considered “ambient” when a model shows negligible performance degradation after reinitializing or randomizing them. 
In contrast, tensors that cause significant performance degradation when altered are termed “important”.  
In other words, certain tensors are deemed “important” because they have a substantial impact on the decision boundary, while others are considered “ambient” due to their smaller effect.
ElasticTrainer~\cite{huang2023elastictrainer} measures the importance of tensors by evaluating the cumulative gradient changes of their weight updates during training. 
This metric aggregates how each weight update contributes to the reduction of training loss, and its computation naturally incorporates the impact of weight dependencies in training.

Directly applying this approach incurs two significant problems.
(1)	Overemphasis on tensors which can be easily trained by the attacker. 
This is because
contribution to the reduction of training loss does not necessarily mean contribution to the model stealing accuracy. 
An important tensor may not necessarily be worth protecting since it can be easily trained by the attacker and its leakage does not lead to a significant increase in model stealing accuracy. In particular, we often find that ElasticTrainer overemphasizes tensors with a large number of parameters. 
(2)	Overemphasis on tensors which be easily acquired from the pre-trained public model. 

To address these problems, we introduce a novel criticalness metric for a DNN tensor $k$ in a specific training epoch as:
\begin{equation}\label{eq:importance}
\footnotesize
  I_k = \underbrace{\sum_{i}  \frac{\partial L }{n  \partial w_i^k } \Delta w_i^k
  }_{\text{Intrinsic tensor importance}} \times \underbrace{\left(1-\cos(f(\mathcal{M}_{Vic}^k), f(\mathcal{M}_{Pub}^k))\right)}_{\text{Attention transition}}, 
\end{equation}
where $L$ denotes the training loss function, $w_i^k$ denotes the $i$-th weight in tensor $k$, 
$n$ is the number of weights in tensor $k$, $\Delta w_i^k$ is the recent update of $w_i^k$ in the training epoch,  $\cos(\cdot)$ is the cosine similarity function, and $f(\cdot)$ is the evaluation function of Grad-Cam~\cite{selvaraju2017grad}. We show the explanations in the following.

\begin{figure}[t]
    \centering
    \includegraphics[width=\linewidth]{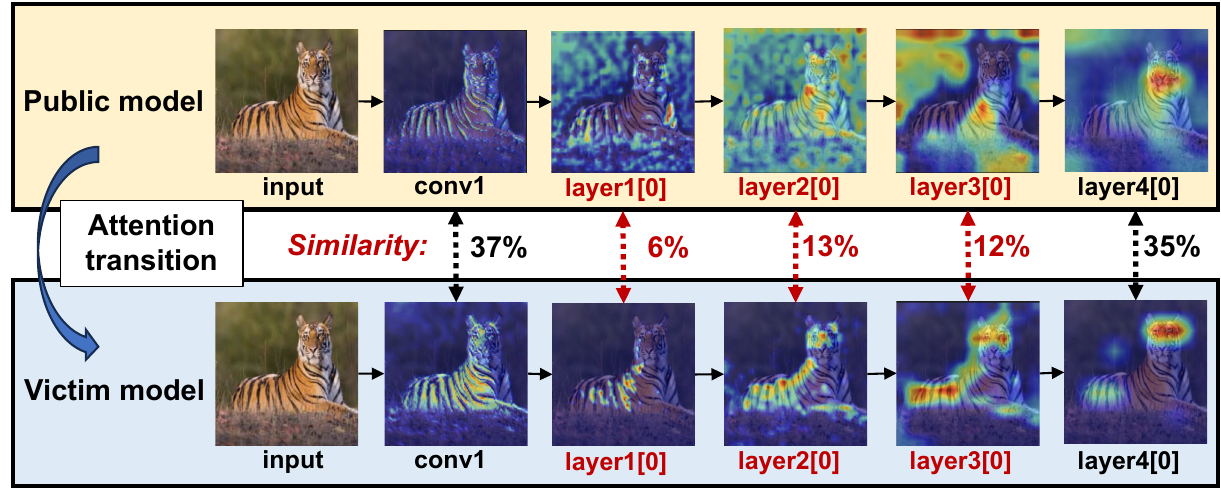}
    \vspace{-2em}
    \caption{An instance of attention transition. Heat maps represent the importance of classifying model decisions.}
    \vspace{-1em}
    \label{fig:grad_cam}
\end{figure}

There are two terms in Eq.~(\ref{eq:importance}):
\begin{itemize}[leftmargin=*]
    \item The first term, intrinsic tensor importance, quantifies the contribution of each weight update to the reduction of training loss, utilizing a gradient computation that mirrors the backward pass, thereby naturally accounting for weight dependencies during training. 
Compared to previous work~\cite{huang2023elastictrainer}, we introduce a new consideration: a regularization term that accounts for the average contribution of parameters within tensors. 
Recent neural network theory~\cite{somepalli2022can} reveals that in the over-parameterized regime, models become highly reproducible, with wide model architectures producing nearly identical decision-making capacities across training runs.
By normalizing the impact of parameters according to their average influence, we address a limitation of previous methods that often overemphasize the importance of parameters solely based on their gradient magnitudes, without considering the collective behavior of parameters within large tensors. 
This regularization provides a more balanced assessment, preventing the overvaluation of tensors with large parameter numbers.
\item The second term, attention transition, quantifies the tensors which can be easily acquired from the pre-trained model. We introduce attention transition into the tensor importance identification.
Our key insight is that if a tensor in the victim model is intrinsically important, but similarly important in the public model, it may not need protection. Specifically, if the public model inherently possesses the capability for a task (e.g., classification decisions), an attacker could learn this capability without needing to steal parameters from the victim model. 
Figure~\ref{fig:grad_cam} shows an example of attention transition using the ResNet18 victim model trained with CIFAR100 dataset and a pre-trained public model. We observe that the attention patterns of the tensor \texttt{layer1.conv0}, \texttt{layer2.conv0}, and \texttt{layer3.conv0} are dissimilar between the victim model and the public pre-trained model, indicating that the victim model has learned new representations during training. 
\end{itemize}

\subsection{Critical Tensor Selection}\label{subsec:design_critical_tensor_selector}
Selecting critical tensors to achieve a pre-defined MS accuracy threshold in a cost-effective selection method is a challenge.
To pinpoint the top-k tensors for protection, we utilize the validation data from the public dataset to simulate the MS attack. 
However, selecting from top-1 to top-k tensors (where 
$k\leq$ \# of tensors) proves time-intensive given the extensive tensor numbers in modern DNNs. 
Once $k$ is set, evaluating the theft precision typically requires extensive training (e.g., 100 epochs \cite{zhang2023no}).

\begin{figure}[t]
    \centering
    \includegraphics[width=\linewidth]{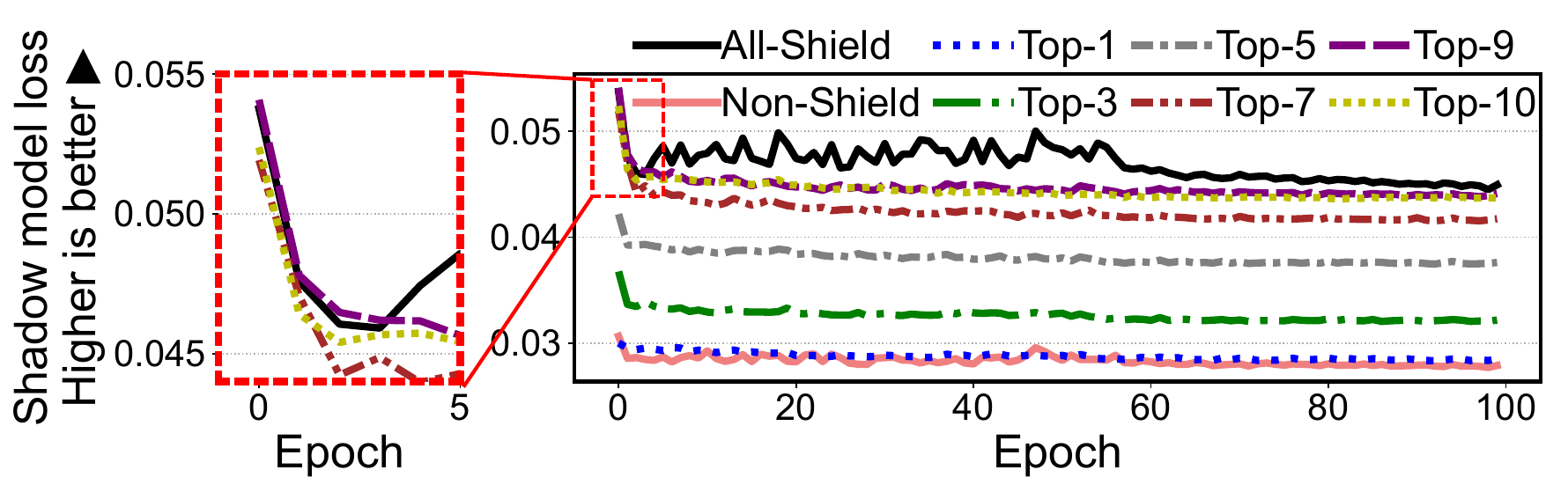}
    \vspace{-2.5em}
    \caption{Model stealing loss values under different top-k strategies. Top-9 protects fewer parameters than Top-10 but also achieves the all-shield protection accuracy.}
    \vspace{-2em}
    \label{fig:loss}
\end{figure}

To design our tensor selection mechanism, we make a fundamental \textit{observation} that is unique in model stealing where the convergence speed of the shadow model's loss function directly reflects the ultimate accuracy of the model stealing.
We illustrate our approach using the VGG16\_BN architecture for both victim and shadow models, leveraging the CIFAR-10 dataset. 
Tensor importance in the victim model is quantified using Equ.~\ref{eq:importance}. 
We protect the top-k layers while using the remaining layers to initialize the shadow model, with attack accuracy depicted in Figure~\ref{fig:loss}.
Our results highlight: (1) Initial loss values provide a coarse measure of protection effectiveness, with initial epochs for top-3 and top-5 layers showing significantly lower values than the All-Shield scenario. 
(2) The rate of loss reduction over the first five epochs offers a fine-grained indicator of protection precision. Specifically, the rate for top-9 and top-10 closely mirrors that of All-Shield, while top-7 exhibits a more gradual decline.

Based on these findings, we propose a critical tensor selection method based on convergence speed. 
The key insight is that instead of waiting for the shadow model to converge, we can assess the accuracy of the top-k shadow model by comparing its convergence speed and initial loss function values with those of the model under All-Shield conditions.
We initiate by evaluating loss variations across $M$ iterations under non-shield and all-shield conditions. 
For each $k$, initial loss values for the corresponding shadow model and the rate of loss reduction over the first 
$m$ epochs (e.g., m=20) are computed. 
If the protection accuracy of the all-shield scenario is met, we methodically decrease the percentage of protected tensors in each layer of the top-k. 
This process is repeated until the protection precision aligns with the threshold of the all-shield scenario.

\section{Critical Feature Identification}\label{subsec:design_selective_masking}
After identifying a set of critical tensors for protection to defend against MS, we have to identify critical intermediate features in order to defend against MIA.

If these intermediate features can be directly observed by the attacker, they can be used to train a membership attack binary predictor (i.e., predicts whether or not the input sample is in the training dataset). 
We say that the intermediate features are critical if their use leads to a high membership inference accuracy. Note that the lower bound of MIA accuracy is 50\%, i.e., attackers can only make random guesses without additional membership privacy information.

The key to preventing privacy leakage of intermediate features is to perform masking so that these features cannot be observed when they are transmitted from TEE to REE. For example, existing work applies one-time padding (OTP) to mask all intermediate features transmitted from the TEE to the REE. The features are then unmasked in the TEE after being transmitted back from the REE to the TEE to recover the original values. 

However, masking all intermediate features imposes significant execution overheads, especially on mobile and IoT devices. 
For example, ShadowNet incurs 160\% and 150\% execution time overheads on the Hikey960 platform for MobileNet and ResNet, respectively, due to the need to refresh masks~\cite{sun2023shadownet}.

Like identifying critical tensors, we would also like to assign each intermediate feature a criticality value to correlate them to the final MIA accuracy. If the intermediate features receive a higher criticality value, they have a larger contribution to the MIA accuracy. After the assignment of criticality values, we can select the top-k intermediate features for masking until we have achieved a predefined threshold of MIA accuracy, e.g., a slight increase compared with masking all intermediate features. 

\begin{figure}[t]
    \centering
    \includegraphics[width=.95\linewidth]{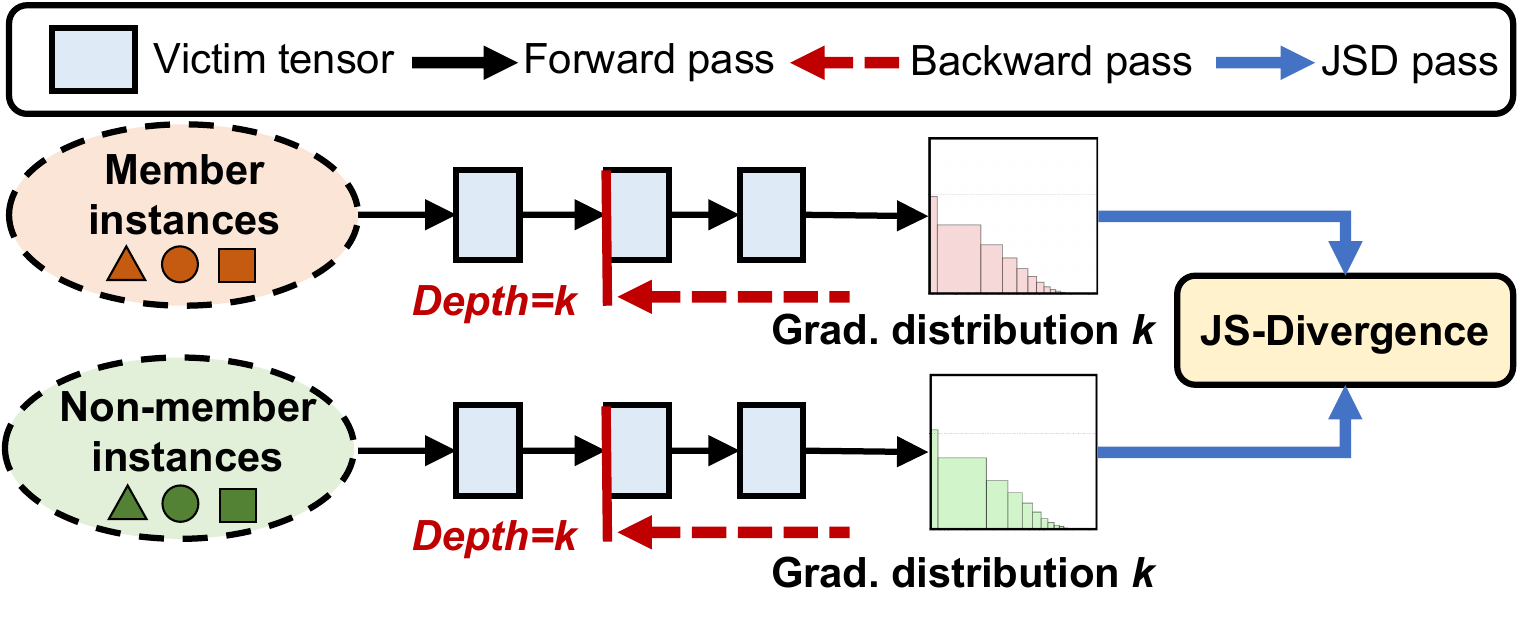}
    \vspace{-1em}
    \caption{Compare the gradient and feature distributions of member and non-member inputs through JS-divergence.}
    \vspace{-1em}
    \label{fig:js_divergence}
\end{figure}
The key insight of our approach is to measure differences the intermediate features exhibit when the target model is given member samples and non-member samples.
This is because the greater the distinction in the internal behavior of the model between member samples (i.e., samples in the victim’s training dataset) and non-member samples (i.e., samples not in the victim’s training dataset), the larger the membership privacy leakage, leading to a higher MIA accuracy. 

We propose a novel membership privacy criticality metric for an intermediate feature, which evaluates membership privacy by calculating the JS-divergence distance of internal information (e.g., gradients) distributions for both members and non-members as:
\begin{equation}\label{eq:JSD}
\footnotesize
\begin{split}
	 JSD(p(z)\Vert q(x))&=\frac{1}{2}KL\left(p\Vert\frac{p+q}{2}\right) +\frac{1}{2}KL\left(q\Vert\frac{p+q}{2}\right),
\end{split}
\end{equation}
where $G_p(x)$ and $G_q(x)$ are gradient (and activation) distributions of member instances and non-member instances, respectively, $KL(\cdot)$ is the Kullback-Leibler(KL) divergence function, and $JSD$ is $\in[0,1]$.
If a pair of distributions is similar, the JSD is near 0; If a pair of distributions is dissimilar, the JSD is near 1.
We select JS-divergence as the distance metric due to its symmetric properties, i.e., $JSD(p(z)||q(x))=JSD(q(x)||p(z))$.
However, the widely used KL-divergence has an asymmetric nature, which does not uniformly characterize the distribution differences between non-member and member samples.

As shown in Figure~\ref{fig:js_divergence}, we assess the membership information exposure of various tensors in the victim model by applying Eq.~(\ref{eq:JSD}) to these intermediate features.

Once the criticality values of all intermediate features are obtained, we can rank them from highest to lowest. A higher score indicates that an intermediate feature is more likely to leak membership privacy. We then select a set of critical features for protection so that the membership inference accuracy does not exceed a given threshold, e.g., $Acc^{AllShield}_{MIA}+1\%$. In our experimental observations, features with a criticality value of less than 0.1 typically do not affect membership inference attack (MIA) accuracy. Therefore, we can set a threshold to filter out features that have negligible distribution distinctions before performing MIA, thereby accelerating the evaluation process.

\section{Latency-aware 
Placement}\label{subsec:design_modeling}
After selecting the critical tensors and intermediate features, TensorShield needs to decide the tensor’s placement based on the hardware performance profile. 

\begin{figure}[t]
    \centering
    \includegraphics[width=\linewidth]{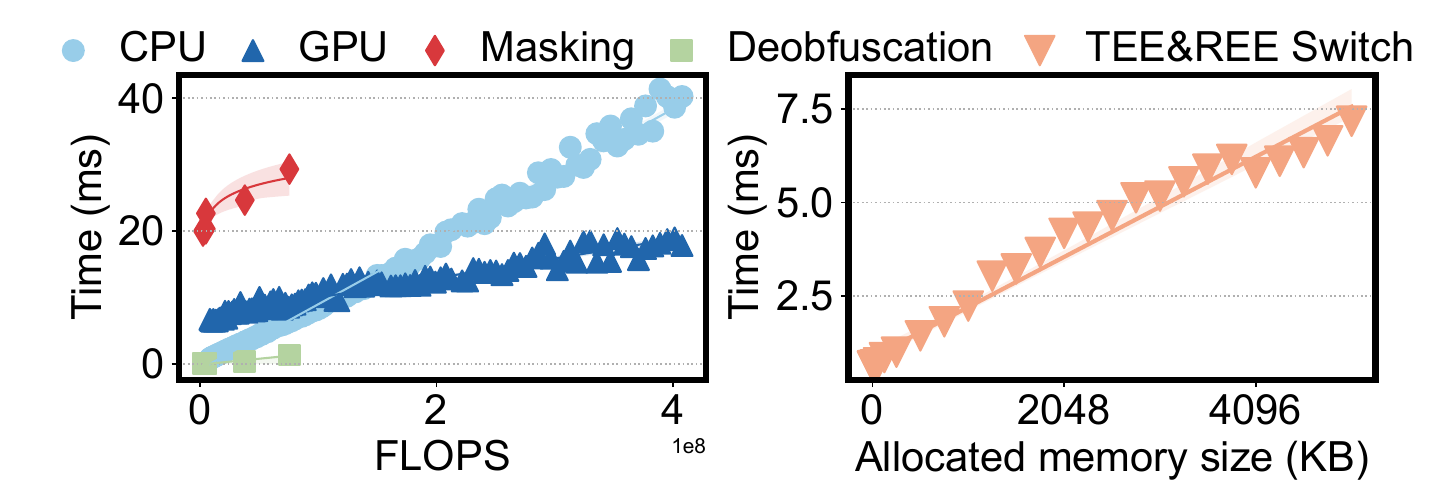}
    \vspace{-1.5em}
    \caption{\textit{Left}: Execution time w.r.t. different FLOPS. \textit{Right}: TEE and REE switching time w.r.t. memory allocation size.}
    \vspace{-2em}
    \label{fig:motivation_computation}
\end{figure}

Due to the unavailability of GPU acceleration within TEE, existing methods~\cite{sun2023shadownet, zhanggroupcover} securely outsource the computation of the model's linear tensor from the TEE to the REE via obfuscation. 
The obfuscated layers can then leverage GPU resources in the REE for accelerated inference.
This approach is effective on powerful platforms (e.g., clouds or PCs), where there is a significant disparity in computing power between CPUs and GPUs. However, our insight is that on mobile and IoT devices, heterogeneous processors exhibit varying affinities for different tensor computations.
We illustrate the execution time for secure operations and various tensors on both CPU and GPU in Figure~\ref{fig:motivation_computation}.
The tests are performed on a Hikey960 mobile device.
We use FLOPS to measure the computational cost of each tensor.
In Figure~\ref{fig:motivation_computation}, we can observe that existing methods overlook two key aspects: (1) Different FLOPS computations are better suited for different accelerators.
For example, smaller FLOPS computations are more efficiently executed on CPUs, while larger FLOPS computations benefit from parallel execution on GPUs. (2) As the communication memory between the REE and TEE increases, the overhead associated with switching increases. 
This drives our effort to comprehensively establish a latency-aware placement method to minimize the inference latency while maintaining model security.

To address the placement issue, we establish a secure inference latency model and formulate it as a numerical optimization problem.
We have introduced the basic placement rules in \S\ref{sec:overview}.
Each tensor is either critical (i.e., requiring shielding) or non-critical. 
Critical tensors can be shielded in two ways: executed directly in the TEE or executed in the REE (after obfuscated offline but still need to be deobfuscated in the TEE online). 
To minimize the Trusted Computing Base (TCB)~\cite{guo2022minimum} size, non-critical tensors are executed directly in the REE. 
We consider there are two types of heterogenous processors, i.e., CPU and GPU, while our formulation could be easily extended to more processors.
There are three options for tensor execution: (1) in the TEE using the CPU; (2) in the REE using the GPU, and (3) in the REE using the CPU.
We use a binary indicator $x_{i,j}$ to denote whether the $i$-th tensor selects the $j$-th execution option in the following:
\begin{equation}
\footnotesize
	 x_{i,j}= \begin{cases}1 & \text {tensor } i \text { is executed with the }j \text{ option} \\ 0 & \text {tensor } i \text { is not executed with the }j \text{ option}\end{cases}.
\end{equation}
We further profile the execution time $t_{i,j}$ of each tensor as follows:
\begin{equation}\label{eq:time}
\footnotesize
     t_{i,j}= \begin{cases}
     t_i^{(REE, CPU)} 
	 & j=0 \\
	 t_i^{(REE, GPU)} 
	 & j=1 \\
	 t_i^{(REE, CPU)}+t_i^{deobf}+t_i^{mask} & j=2 \\
	 t_i^{(REE, GPU)}+t_i^{deobf}+t_i^{mask} & j=3 \\ 
	 t_i^{(TEE, CPU)} & j=4 \\ 
	 \end{cases},
\end{equation}
where $t_i^{deobf}$ and $t_i^{mask}$ denote the deobfuscation time for tensors and masking (including unmasking) for privacy-related features, respectively.
Our optimization objective is to minimize the total inference time, which is determined by the computation time of each tensor and TEE and REE switching time:
\begin{equation}\label{eq:objective}
\footnotesize
\begin{split}
\underset{x}{\arg\min} &\underbrace{\sum_{i\in N, j\in{0,1,2}}x_{i,j}\cdot t_{i,j}}_{\text{computation time}}+\underbrace{\left(\sum_{i\in{N-1}}x_{(i-1),0}\oplus (\sum_{j=1}^4 x_{i, j})\right)\cdot t^{switch}}_{\text{switch time}}
\\
s.t. &\quad \text{(Correctness)}\quad \sum_{j\in{0,1,2}} x_{i,j}=1,\forall_{i\in N}
\\
&\quad \text{(Memory)}\quad\quad\,\,\, x_{i,0}\cdot m_i\leq M, \forall_{i\in N}
\end{split}
\end{equation}
where $t^{switch}$ denotes the switch time, $m_i$ denotes the allocated TEE memory of the $i$th tensor, and $M$ denotes a specified TEE available memory.
Note that a switch in execution environments occurs only if the current tensor and the preceding tensor are processed in different execution environments. 
To model this, we use the $\oplus$ operation to represent the switch.

Our problem is related to the traveling salesman problem and is NP-hard~\cite{xing2008rendezvous}. 
Fortunately, we have observed that the proportion of time spent on switching environments is trivial compared to the execution time, allowing us to disregard the overhead associated with environment switching. 
Consequently, by relaxing the condition, we can decompose the problem into multiple subproblems, enabling us to derive an approximately optimal solution with mature solvers:
\begin{equation}\label{eq:solution}
\footnotesize
	 x_{i,j}= \begin{cases}1& j=j_i^{opt}=\begin{cases}\underset{j\in{0,1}}{\arg\min}\{ t_{i,j}\}&m_i\leq M\\\underset{j\in{1,2}}{\arg\min}\{ t_{i,j}\}&m_i> M
	 \end{cases} \\ 0 & j\ne j_i^{opt}\end{cases}.
\end{equation}

\section{Implementation}
We implement the \name~as an end-to-end system consisting of two parts: offline profiling and on-device inference.
In offline profiling, we leverage the Knockoff Net~\cite{orekondy2019knockoff} as an emulator for MS to identify critical tensors. 
It is a standard query-based stealing technique where the attacker trains a model from a set of collected data labeled by the $\mathcal{M}_{Vic}$~\cite{jagielski2020high, shen2022model}.
For MIA, we employ a transfer attack that builds $\mathcal{M}_{Sur}$ to imitate the behavior of $\mathcal{M}_{Vic}$ and infer the privacy of $\mathcal{M}_{Vic}$ from white-box information of $\mathcal{M}_{Sur}$~\cite{salem2018ml}.
We leverage the attack implementation from a recent benchmark suite, ML-\textsc{Doctor}~\cite{liu2022ml, shen2022model}.
We conduct victim and surrogate model training using PyTorch 2.1.
To support on-device secure inference, we adopt DarkneTZ~\cite{mo2020darknetz}, the state-of-the-art on-device secure inference framework.
We add $\sim$2.4K LOC in C to DarkneTZ's TEE modules and $\sim$3.2K LOC in C to  DarkneTZ's REE modules for deobfuscating, masking, and unmasking in the TEE.
Our code is open-sourced for public access\footnote{\url{https://github.com/suntong30/TensorShield}}.

\begin{table}[t]
\caption{Device specifications in our evaluation.}
\vspace{-1em}
\label{tab:evaluated_platform}
\resizebox{\linewidth}{!}{
\begin{tabular}{ccccccc}
\toprule
\textbf{Device Name }    & \textbf{CPUs}                                                                                          & \textbf{GPU}                                                         & \textbf{RAM} & \begin{tabular}[c]{@{}c@{}}\textbf{TEE}\\ \textbf{RAM}\end{tabular} & \textbf{REE OS}      &\textbf{TEE OS} \\ \toprule
Hikey960~\cite{hikey960}        & \begin{tabular}[c]{@{}c@{}}4xCortex-A73\\ (@2.36 GHz)\\ 4xCortex-A53\\ (@1.84 GHz)\end{tabular} & \begin{tabular}[c]{@{}c@{}}ARM Mali \\ G71 MP8\end{tabular} & 4 GB & 64 MB                                              & Android 7   & \tabincell{c}{OP-TEE \\ v3.4.0}  \\ \hline
\tabincell{c}{Raspberry Pi 3B+ \\ (RPI3B+)}~\cite{rpi3b+} & \begin{tabular}[c]{@{}c@{}}4xCortex-A53\\ (@1.40 GHz)\end{tabular}                             & \tabincell{c}{VideoCore 4}                                                & 1 GB & 32 MB                                              & Raspbian OS & \tabincell{c}{OP-TEE \\ v3.4.0} \\ \bottomrule
\end{tabular}
}
\vspace{-1em}
\end{table}


\begin{table}[t]
\centering
\caption{Accuracy of $\mathcal{M}_{vic}$ used in the evaluation.}
\label{tab:eva_models}
\vspace{-1em}
\resizebox{\linewidth}{!}{
\begin{tabular}{ccccc}
\toprule
            & \textbf{CIFAR-10}~\cite{krizhevsky2009learning}         & \textbf{CIFAR-100}~\cite{krizhevsky2009learning}& \textbf{STL-10}~\cite{coates2011analysis}           & \textbf{Tiny-ImageNet}~\cite{le2015tiny} \\ 
\midrule
ResNet18~\cite{he2016deep}    & 86.71\% & 60.72\% &         86.73\%        & 42.96\% \\ 
MobileNetV2~\cite{sandler2018mobilenetv2} & 83.17\% & 50.06\% & 74.36\% & 41.26\% \\ 
VGG16\_BN~\cite{simonyan2014very}   & 85.28\% & 71.59\% &        80.20\%         & 41.28\% \\ 
ResNet50~\cite{he2016deep}    & 86.47\%         & 63.95\%         & 87.28\%         & 54.45\% \\ 
\bottomrule
\end{tabular}
}
\end{table}

\section{Evaluation}\label{sec:evaluation}
The key takeaways of our evaluation are:
\begin{itemize}[leftmargin=*]
    \item Across four datasets and four models, \name~can reduce inference latency and energy consumption by up to 25.35$\times$ (avg. 5.85$\times$) and up to 91.35\% (avg. 58.66\%) compared to the state-of-the-art obfuscation baseline~\cite{zhanggroupcover}, and up to 16.89$\times$ (avg. 4.32$\times$) and up to 98.35\% (avg. 69.86\%) compared to the state-of-the-art shielding the whole models baseline~\cite{lee2019occlumency}, while maintaining almost same MS (1.03$\times$) and MIA (1.00$\times$) accuracy.
    \item \name's critical tensor evaluator can reduce selected parameters by average relatively 70.32\% (absolutely 17.62\%) compared to the state-of-the-art XAI-based evaluating method~\cite{huang2023elastictrainer} for achieving the same black-box MS and MIA accuracy.
    \item \name's improvement is significant across various workloads and platforms, including different input sizes, model architectures, and hardware devices. 
    \item \name's system overhead is negligible.
\end{itemize}

\subsection{Experimental Setup}\label{subsec:eva_setup}
\textbf{Platforms.}
We conduct the evaluations on two types of mobile and IoT devices with different hardware specifications, all equipped with Armv8-A CPUs and ARM TrustZone.
The detailed specifications are listed in Table~\ref{tab:evaluated_platform}.
Hikey960 is a mobile development board equipped with HiSilicon Kirin960 SoC, and Raspberry Pi 3B+ (RPI3B+) is a typical IoT device. 
We follow the literature~\cite{sun2023shadownet} to expand the TEE RAM (i.e., secure memory).

\textbf{Models and datasets.}
We evaluate \name~with 4 typical DNN models that are widely used for device learning (i.e., MobileNetV2~\cite{sandler2018mobilenetv2}, ResNet-18~\cite{he2016deep}, ResNet-50~\cite{he2016deep}, VGG16\_BN~\cite{simonyan2014very}).
We use 4 popular datasets, i.e., CIFAR-10~\cite{krizhevsky2009learning}, CIFAR-100~\cite{krizhevsky2009learning}, STL-10~\cite{coates2011analysis}, and Tiny-ImageNet~\cite{le2015tiny} datasets with resized input size 3$\times$32$\times$32, 3$\times$32$\times$32, 3$\times$128$\times$128, and 3$\times$224\allowbreak$\times$224. 
The MobileNetV2 is the simplest model and the VGG16\_BN is the most complex model in our evaluation.
The CIFAR-10 and STL-10 are simple datasets while CIFAR-100 and Tiny-ImageNet are complex datasets.
The dataset and model
selection refers to prior secure inference literatures~\cite{liu2022ml,orekondy2019knockoff,zhang2023no,sun2023shadownet,mo2020darknetz,zhanggroupcover}.

\textbf{Baselines.}
We compare the performance of \name\ with the following 7 baselines, consisting of 1 no-shield (i.e., Native), 4 partial-shield (i.e., DarkneTZ~\cite{mo2020darknetz}, Serdab~\cite{elgamal2020serdab}, Magnitude~\cite{hou2021model}, and ShadowNet~\cite{sun2023shadownet}), and 2 all-shield (i.e., GroupCover~\cite{zhanggroupcover} and Occlumency~\cite{lee2019occlumency}) solutions:
\begin{itemize}[leftmargin=*]
    \item \textbf{Native} executes the whole victim model in REE.
    \item \textbf{DarkneTZ}~\cite{mo2020darknetz} shields victim model's deep layers.
    \item \textbf{Serdab}~\cite{elgamal2020serdab} shiedls victim model's shallow layers.
    \item \textbf{Magnitude}~\cite{hou2021model} shields victim model's large magnitude weights.
    \item \textbf{ShadowNet}~\cite{sun2023shadownet} shields non-linear layers and obfuscates linear layers in a fixed strategy.
    \item \textbf{GroupCover}~\cite{zhanggroupcover} shields non-linear layers and obfuscates linear layers in a random strategy.
    \item \textbf{Occlumency}~\cite{lee2019occlumency} shields all victim weights in TEE.
\end{itemize}

\textbf{Metrics.}
There are two types of metrics in our evaluations.
(1) Security metrics.
We use the MS accuracy and MIA accuracy as model security metrics.
The MS accuracy measures how many test samples can be correctly classified by the attacker's surrogate model.
Achieving high accuracy is a primary goal of model stealing attacks.
The MIA accuracy represents the membership classification accuracy.
(2) System metrics. 
For inference latency, we use the model inference time as our main metric which
measures the time between feeding an input and getting the output.
We also measure the energy consumption during the inference.

\textbf{Configuration.}
For all cases, we use the public models as initialization to get a better model performance.
We follow the hyper-parameter settings of Knockoff Nets~\cite{orekondy2019knockoff}.
We use a minibatch size of 64, select cross-entropy loss, use SGD with weight decay of 5e-4, a momentum of 0.5, and train the victim models for 100 epochs.
The learning rate is originally set to 0.1 and decays by 0.1 every 60 epochs.
For the training in the MIA part, we follow the settings
of \textsc{ML-Doctor}.
The learning rate is 1e-2.
The hyper-parameters to train shadow models are the same as victim models.
We set the JS-divergence threshold as 0.1, a common value for evaluating two distributions~\cite{deasy2020constraining, li2022multi}.
To ensure that the
results on the device are not impacted by the throttle, we lock the CPU frequency and set sufficient time gaps between each test.
We run all experiments ten times and record the average inference time.

\subsection{Overall performance}
\begin{figure*}[t]
    \centering
    \includegraphics[width=\linewidth]{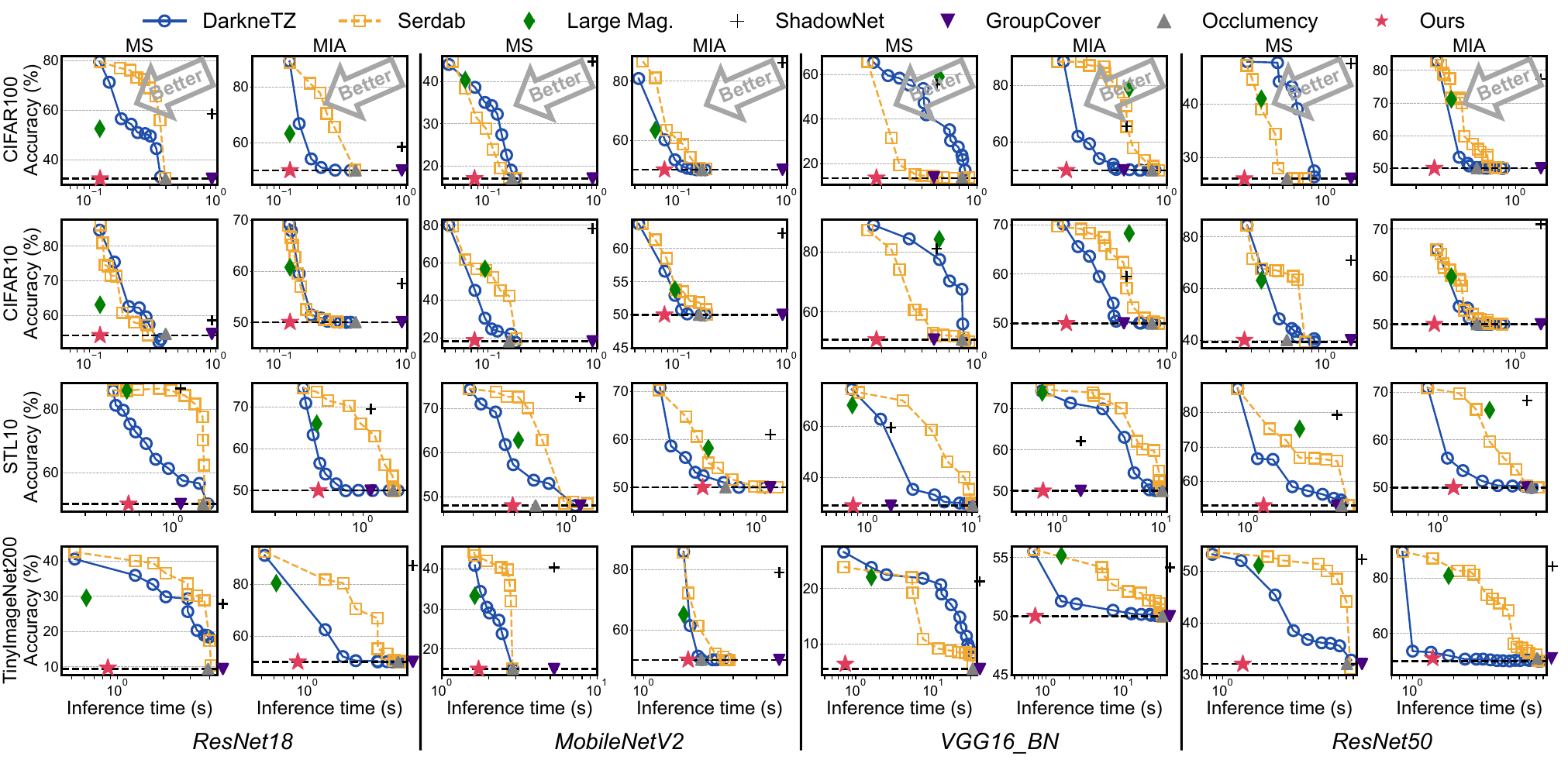}
    \vspace{-2em}
    \caption{Model stealing (MS) and membership inference attack (MIA) accuracy and inference time regarding baselines on Hikey960.
    The dotted line ``- - -'' represents black-box accuracy.
    The points of GroupCover and ShadowNet outside the right border of the subfigure represent OOM (i.e., out-of-memory error in TEE). \name~(Ours) shields critical tensors, achieving inference speeds up to 25.35$\times$ (avg. 5.85$\times$) and 16.89$\times$ (avg. 4.32$\times$) faster than GroupCover~\cite{zhanggroupcover} and Occlumency~\cite{lee2019occlumency}, respectively.}
    \vspace{-1em}
    \label{fig:MS_MIA_Latency}
\end{figure*}
Figure~\ref{fig:MS_MIA_Latency} shows the overall results comparing \name~with all baselines.
We follow the attack pipelines in \S\ref{sec:background}.
For DarkneTZ~\cite{mo2020darknetz} and Serdab~\cite{elgamal2020serdab}, we use the offloaded layers to replace the corresponding layers of $\mathcal{M}_{init}$.
For Magnitude~\cite{hou2021model}, we use the offloaded
weights to replace the corresponding
weights in $\mathcal{M}_{init}$. 
For ShadowNet~\cite{sun2023shadownet}, the attacker uses the
public model to decode the obfuscation algorithm and uses the decoded weights to initialize $\mathcal{M}_{init}$.
The accuracy of $\mathcal{M}_{vic}$ used in the evaluation is shown in Table~\ref{tab:eva_models}.

As shown in Figure~\ref{fig:MS_MIA_Latency}, \name~achieves an inference speedup of up to 25.35$\times$ (avg. 5.85$\times$) and up to 16.89$\times$ (avg. 4.32$\times$) compared to GroupCover and Occlumency, respectively.
Meanwhile, \name~achieves the same security protection as shielding the entire model inside TEE, i.e., 1.03$\times$ for MS and 1.00$\times$ for MIA.
We can make the following observations:

(i) Under the same model architecture, when compared with shielding partial layers (i.e., Serdab and DarkneTZ) that achieve nearly identical inference latency as TensorShield, TensorShield offers better protection against MS in more complex datasets. For example, the first and last rows show better MS protection results than the second and third rows in Figure X. This is because complex datasets require both shallow and deep tensors in DNNs to perform comprehensive feature extraction, a task not achievable by tensors fixed at certain depths. TensorShield is able to identify critical tensors at varying depths to defend against MS attacks.

(ii) Within the same dataset, as model complexity increases, TensorShield provides better protection against MIA compared to shielding partial layers (i.e., Serdab and DarkneTZ) that achieve nearly identical inference latency as TensorShield. For example, for the STL10 dataset, TensorShield’s effectiveness in protecting against MIA increases for MobileNetV2, ResNet18, ResNet50, and VGG16\_BN. This is because as the complexity of models increases, it becomes evident that not only the final classifier tensor harbors substantial membership privacy information, but the preceding tensors do as well. This indicates that for simple models, protecting only the final classifier layer is often sufficient. However, for more complex models, such protection is inadequate to satisfy security requirements against MIA.

(iii) Compared to shielding all layers (i.e., ShadowNet, GroupCover, and Occlumency), the more complex the model, the greater the improvement in inference performance by TensorShield. 
For instance, on the TinyImage200 dataset compared to Occlumency, TensorShield shows substantial performance improvements for VGG16\_BN, ResNet50, ResNet18, and MobileNetV2 by 16.89$\times$, 5.78$\times$, 4.54$\times$, and 1.66$\times$, respectively. 
This is because simpler models are more computationally suited to CPU processing, whereas more complex models are better handled by TensorShield, which could execute critical tensors in the REE by obfuscating them on the GPU.

(iv) The shapes of the curves for shielding deep layers (i.e., DarkneTZ) and shallow layers (i.e., Serdab) differ significantly across datasets and models. For a given victim model, setting a uniform threshold is challenging without comprehensive empirical measurements of both security and inference efficiency. This variability arises because the ``sweet spots'' for optimal shielding differ across datasets. 
For example, when shielding deep layers of ResNet18 to prevent model stealing (the first column in Figure X ), the curve shapes for CIFAR-10 and CIFAR-100 differ markedly from those for STL-10. 

(v) As the complexity of a dataset increases, the attack accuracy of MIA tends to rise while the model’s MS accuracy tends to decrease. This phenomenon can be attributed to the increased difficulty in comprehending the model’s decision-making processes when faced with complex datasets, which also heightens the propensity for overfitting. Such overfitting characterized by a lack of generalization, results in a more pronounced leakage of membership information. 

\begin{figure*}[t]
    \centering
    \includegraphics[width=\linewidth]{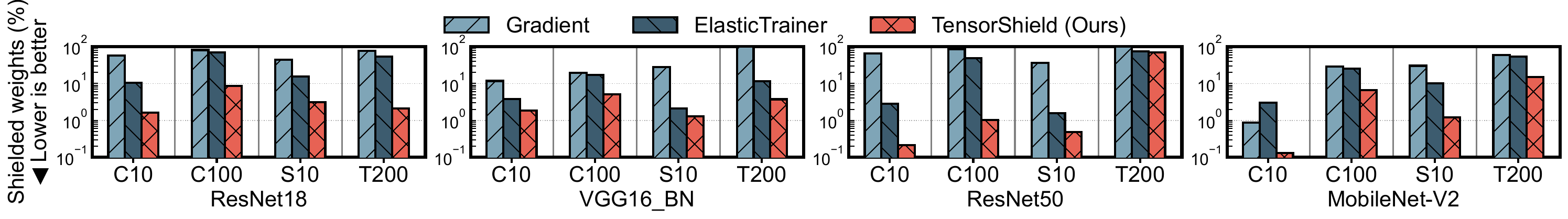}
    \vspace{-1.5em}
    \caption{Comparison of representative XAI-based tensor selection methods.}
    \vspace{0em}
    \label{fig:eva_XAI_weights}
\end{figure*}
\textbf{Understanding \name's improvements.}
TensorShield outperforms various baselines for slightly different reasons.
(i) Compared to the shielding none or partial layers (or weights) baselines, \name~has lower MS and MIA accuracy that achieves all-shield protection efficacy, because it dynamically selects critical tensors based on our critical tensor identification method (cf., \S\ref{sec:critical_tensor_identification}) for different datasets and model architectures instead of fixed selection solutions.
We will show the efficacy of our critical tensor evaluator in \S\ref{subsec:eva_critical_tensor_evaluator}.
Since fixed obfuscation is easy to reverse and restore victim weights, compared to ShadowNet, TensorShield still achieves higher model security. 
(ii) Compared to the all-shield baselines, \name~only masks privacy-related features (cf., \S\ref{subsec:design_selective_masking}) and plans tensor optimal execution strategy based on computation modeling (cf., \S\ref{subsec:design_modeling}), thus being able to reduce the inference latency.  
We will show details of runtime performance in \S\ref{subsec:eva_runtime}.




\subsection{Efficacy of Critical Tensor Identification}\label{subsec:eva_critical_tensor_evaluator}
Next, we show the efficacy of critical tensor identification (cf. \S\ref{sec:critical_tensor_identification}).
We first compare \name~with two representative XAI methods \cite{sundararajan2017axiomatic,huang2023elastictrainer}: the Gradient method~\cite{sundararajan2017axiomatic} utilizes integrated gradients, whereas ElasticTrainer~\cite{huang2023elastictrainer} represents state-of-the-art work combining both gradients and weight updates.
To ensure a fair comparison, we evaluate the importance of tensors using these methods and then rank their importance values. 
Tensors are selected for protection from highest to lowest until they meet the specified MS accuracy threshold (i.e., $Acc^{All}_{MS} + 1\%$).
As shown in Figure~\ref{fig:eva_XAI_weights},
\name~selects average 87.82\% (absolutely 50.60\%) fewer parameters compared to the Gradient method, and relatively 70.32\% (absolutely 17.62\%) fewer than ElasticTrainer. 
Notably, \name~exhibits superior performance on datasets with fewer classification tasks (e.g., CIFAR-10 and STL-10).
This improvement can be attributed to the public pre-trained models carrying more decision-making information from simpler datasets. 
\name~effectively reduces the importance of these tensors, thus excluding them from selection (cf. \S\ref{subsec:design_critical_tensor}).
\begin{figure}[t]
    \centering
    \includegraphics[width=\linewidth]{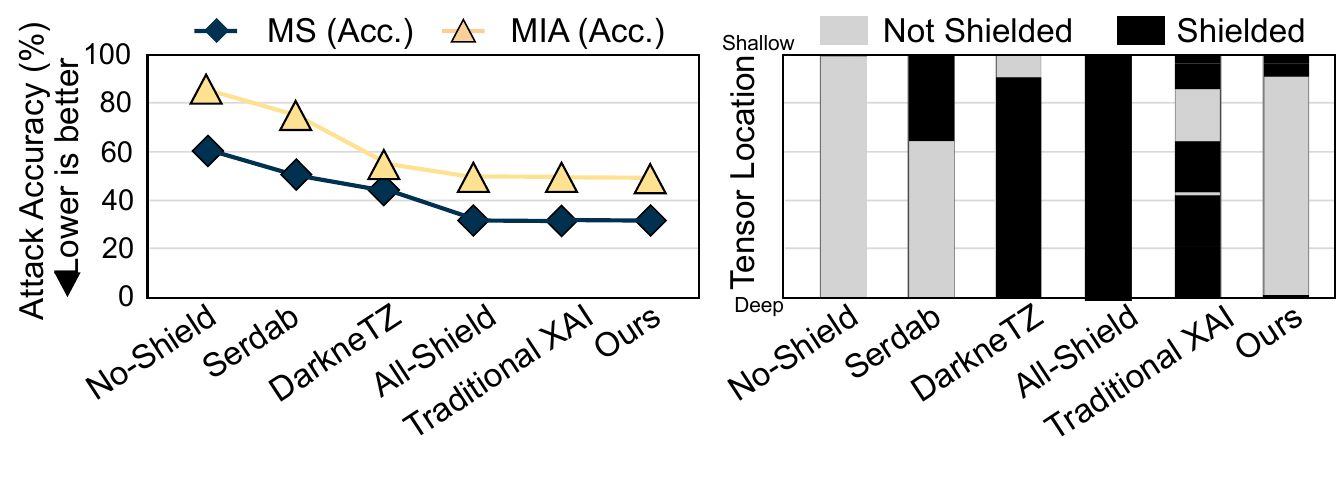}
    \vspace{-2.5em}
    \caption{Attack accuracy (lines) regarding different defense schemes (bars) for ResNet18 with CIFAR100 dataset. 
    We can only shield $\sim$8\% weights (we show the selected tensors in Figure~\ref{fig:critical_tensor}) within TEE to achieve comparable protection efficacy as shielding all weights.}
     \vspace{-2em}
    \label{fig:motivation_selection}
\end{figure}

\begin{figure}[t]
    \centering
    \includegraphics[width=\linewidth]{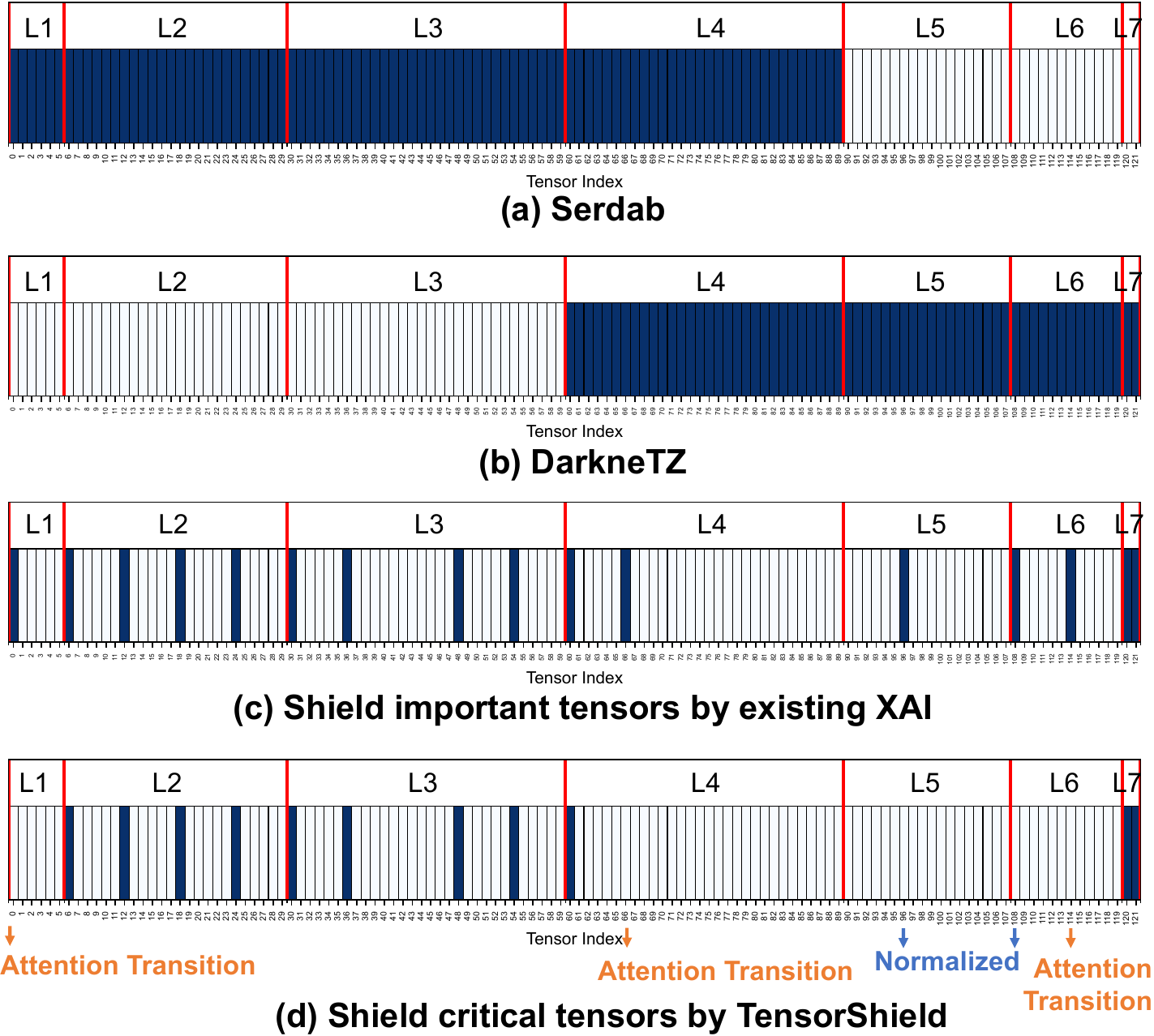}
    \vspace{-1.5em}
    \caption{Tensor selections regarding different defense schemes for ResNet18 with CIFAR100 dataset. (a) Serdab: shields four shallow layers. (b) DarkneTZ: shields four deep layers. (c) Shielding important tensors by traditional XAI (i.e., ElasticTrainer). (d) \name: shields critical tensors. The parameters are shown in Figure~\ref{fig:motivation_selection}.}
    \label{fig:critical_tensor}
\end{figure}

To better understand the efficacy, we present in Figure~\ref{fig:motivation_selection} the MS accuracy of a ResNet18 model trained on the CIFAR100 dataset using different defense methods. 
We configure Serdab and DarkNeTZ to protect an identical number of layers (i.e., four layers). 
In contrast, traditional XAI methodologies employ ElasticTrainer to safeguard critical tensors. Our key observations include: (i) Although DarkNeTZ protects over 85\% of the model parameters, it fails to secure the model effectively. 
In comparison, TensorShield adeptly identifies critical tensors distributed across both shallow and deep tensors in the victim model. 
(ii) Although finding important tensors via traditional XAI can achieve model security, the volume of parameters it selects far exceeds that of TensorShield. 
This is because important tensors are not necessarily critical tensors. 
The tensors selected by each method are illustrated in Figure~\ref{fig:critical_tensor}. 
Notably, TensorShield (Figure~\ref{fig:critical_tensor}(d)) excludes five important tensors compared to ElasticTrainer (Figure~\ref{fig:critical_tensor}(c)).
Specifically, TensorShield excludes three tensors because they perform as well as those in the pre-trained public model (cf. the attention transition term in Eq.~(\ref{eq:importance})). It also excludes two tensors due to their excessive parameter count, as determined by the intrinsic importance normalization term in Eq.~(\ref{eq:importance}).
Additionally, the order of tensor criticality in TensorShield significantly diverges from the importance ranking used by ElasticTrainer, thereby demonstrating the superior efficacy of TensorShield’s critical tensor identification mechanism.

\begin{figure}[t]
    \centering
    \includegraphics[width=\linewidth]{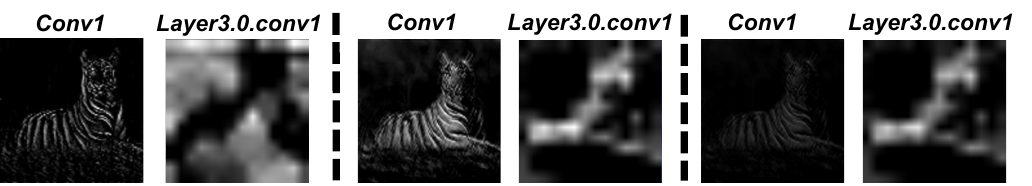}
    \vspace{-2em}
    \caption{Visualization of critical tensor evaluation. Higher brightness indicates greater attention. \textit{Left}: Public model's attention. \textit{Middle}: Victim model's attention. \textit{Right}: Critical tensor evaluation results.}
    \vspace{0em}
    \label{fig:eva_attention}
\end{figure}

We visualize \name's critical tensor evaluation process in Figure~\ref{fig:eva_attention}.
The decision-making attention of the \texttt{conv1} tensor in the victim model is similar to that in the public model, whereas \texttt{layer3.0.conv1} exhibits significant differences. 
This indicates that \texttt{layer3.0.conv1} contains more critical information specific to the victim model. Therefore, as shown in Figure~\ref{fig:eva_attention}(right), \name~reduces the importance of the \texttt{conv1} tensor while preserving the significance of \texttt{layer3.0.conv1}.

\subsection{Runtime Performance}\label{subsec:eva_runtime}
\textbf{Overall latency.}
\begin{figure*}[t]
    \centering
    \begin{minipage}[t]{\linewidth}
    \centering
    \includegraphics[width=.95\linewidth]{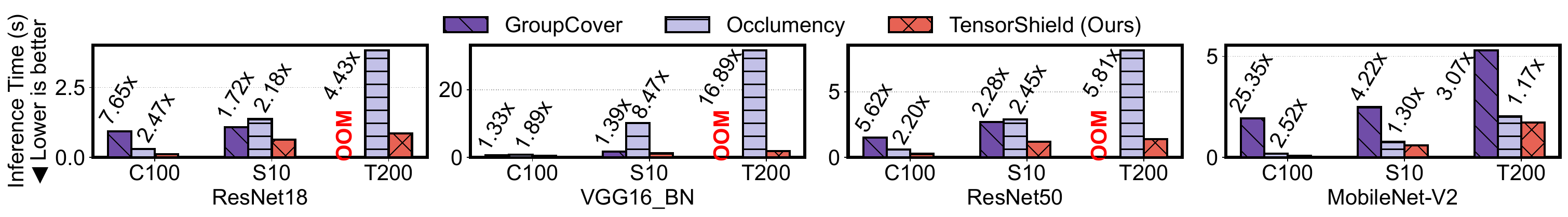}
    \end{minipage}
    \vspace{-1em}
    \begin{minipage}[t]{\linewidth}
    \centering
    \includegraphics[width=.95\linewidth]{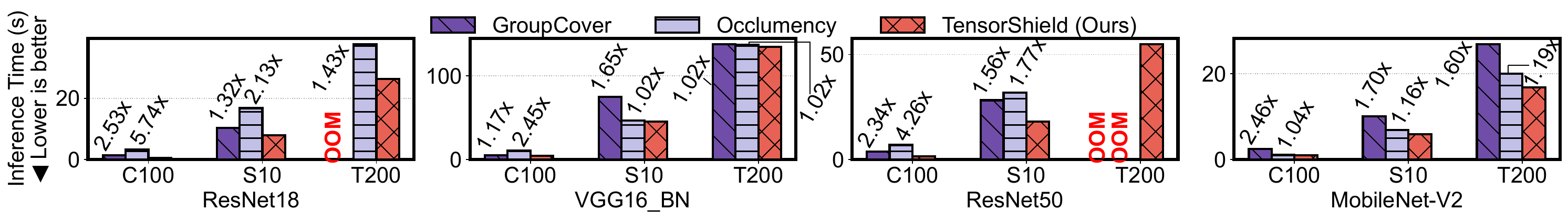}
    \end{minipage}
    \caption{Inference time of 4 models compared with native and two completely secure baselines. \textit{Top:} Hikey960. \textit{Bottom:} RaspberryPi 3B+.}
    \label{fig:eva_inference_time}
\end{figure*}
We conduct on-device inference experiments on the four models using three datasets (i.e., CIFAR-100, STL10, and Tiny-ImageNet).
To ensure fairness, we compare \name~with baselines that achieve the black-box protection security effects (i.e., GroupCover and Occlumency).
The results on two hardware platforms are shown in Figure~\ref{fig:eva_inference_time}.
On mobile devices (i.e., Hikey960), \name's inference time is up to 25.35$\times$ (avg. 5.85$\times$) and 16.89$\times$ (avg. 4.32$\times$) faster than GroupCover and Occlumency, respectively. 
This latency speedup is due to \name's computation power modeling and the reduced overhead from masking, which we will detail in the following breakdown experiment. Even on IoT devices lacking GPU acceleration (i.e., Raspberry Pi 3B+), \name~achieves inference times that are up to 2.53$\times$ (avg. 1.74$\times$) and 5.74$\times$ (avg. 2.11$\times$) faster than GroupCover and Occlumency, respectively.

\textbf{Breakdown.}
To better understand the reasons behind \name's speedup, we conduct a breakdown of inference times on the Hikey960, as shown in Table~\ref{tab:eva_breakdown}.
The inference process for GroupCover includes calculation, communication, masking, and deobfuscation, while Occlumency’s process involves only calculation and decryption. 
We make two observations: (i) Depending on the size of the input, \name~adaptively chooses whether to execute tensors within the TEE or after obfuscating them to the REE. 
For example, for the CIFAR-100 dataset, due to the smaller computational demand, executing critical tensors in the TEE using the CPU is faster than after obfuscation on the GPU. 
Therefore, \name~computes critical tensors in the TEE without needing to mask intermediate features, whereas GroupCover obfuscates all parameters to the REE GPU and masks all intermediate features. 
Compared to Occlumency, \name~saves on the additional overhead of tensor computations and decryption in the TEE. 
For the STL-10 and Tiny-ImageNet datasets, which require heavier computations, GPU execution significantly outpaces CPU, hence \name~obfuscates tensors to the REE for GPU-accelerated execution, achieving a 3.71$\times$-9.82$\times$ improvement in calculation time over Occlumency.
(ii) \name~only masks the features of the last two tensors in the STL-10 dataset, reducing the masking time by 52.60\% compared to GroupCover.
This is attributed to the membership-aware masking technique.

\begin{table}[t]
\caption{Inference time breakdown for ResNet18 model with CIFAR-100 dataset. "Cal", "Dec", "Comm", "Mask", and "Deobf" refer to calculation, decryption, communication, masking, and deobfuscation. \ding{55} represents OOM.}
\label{tab:eva_breakdown}
\vspace{-1em}
\resizebox{\linewidth}{!}{
\begin{tabular}{c|c|cccccc}
\Xhline{1pt}
\multirow{2}{*}{\textbf{Dataset}}   & \multirow{2}{*}{\textbf{Work}}                    & \multicolumn{6}{c}{\textbf{Execution time}}                                                                                                                             \\ \cline{3-8} 
                         &                                          & \textbf{Cal.}                 & \textbf{Dec.}             & \textbf{Comm.}                & \textbf{Mask.}              & \textbf{Deobf.}                          & \textbf{Total}                \\ \hline
                      \hline
\multirow{3}{*}{\STAB{\rotatebox[origin=c]{90}{C100}}}   & Occlumency~\cite{lee2019occlumency}                               &      0.219s                &      0.080s                &      /                &      /                &     /                 &       0.299s                  \\ \cline{2-8} 
                         & GroupCover~\cite{zhanggroupcover}                               &      0.497s                &        /              &           0.023s           &      0.403s                &        0.049s              &            0.926s                      \\ \cline{2-8} 
                         & \multicolumn{1}{c|}{\textbf{Ours}} &  \multicolumn{1}{c}{\textbf{0.122s}} &  \multicolumn{1}{c}{/} & \multicolumn{1}{c}{\textbf{0.002s}} & \multicolumn{1}{c}{/} & \multicolumn{1}{c}{/} & \multicolumn{1}{c}{\textbf{0.124s}}  \\ \hline
\multirow{3}{*}{\STAB{\rotatebox[origin=c]{90}{S10}}} & Occlumency~\cite{lee2019occlumency}                               &        1.288s              &          0.080s            &              /        &          /            &  /                    &       1.368s                    \\ \cline{2-8} 
                         & GroupCover~\cite{zhanggroupcover}                               &     0.510s                 &           /          &    0.065s                   &         0.481s             &         0.027s             &    1.083s                                   \\ \cline{2-8} 
                         & \multicolumn{1}{c|}{\textbf{Ours}} & \multicolumn{1}{l}{\textbf{0.347s}} & \multicolumn{1}{c}{/} & \multicolumn{1}{l}{\textbf{0.037s}} & \multicolumn{1}{l}{\textbf{0.228s}} & \multicolumn{1}{l}{\textbf{0.016s}} & \multicolumn{1}{l}{\textbf{0.628s}}  \\
                         \hline
\multirow{3}{*}{\STAB{\rotatebox[origin=c]{90}{T200}}} & Occlumency~\cite{lee2019occlumency}                               &        3.741s              &          0.080s            &              /        &          /            &  /                    &       3.821s                    \\ \cline{2-8} 
                         & GroupCover~\cite{zhanggroupcover}                               &     \ding{55}                 &      \ding{55}          &    \ding{55}                  &        \ding{55}            &        \ding{55}          &    \ding{55}                                   \\ \cline{2-8} 
                         & \multicolumn{1}{c|}{\textbf{Ours}} & \multicolumn{1}{l}{\textbf{0.381s}} & \multicolumn{1}{c}{/} & \multicolumn{1}{l}{\textbf{0.056s}} & \multicolumn{1}{l}{\textbf{0.364s}} & \multicolumn{1}{l}{\textbf{0.062s}} & \multicolumn{1}{l}{\textbf{0.863s}}  \\
                         \Xhline{1pt}
\end{tabular}
}
\end{table}
\subsection{Energy Consumption}
\begin{figure}[t]
    \centering
    \begin{minipage}[t]{\linewidth}
    \centering
    \includegraphics[width=\linewidth]{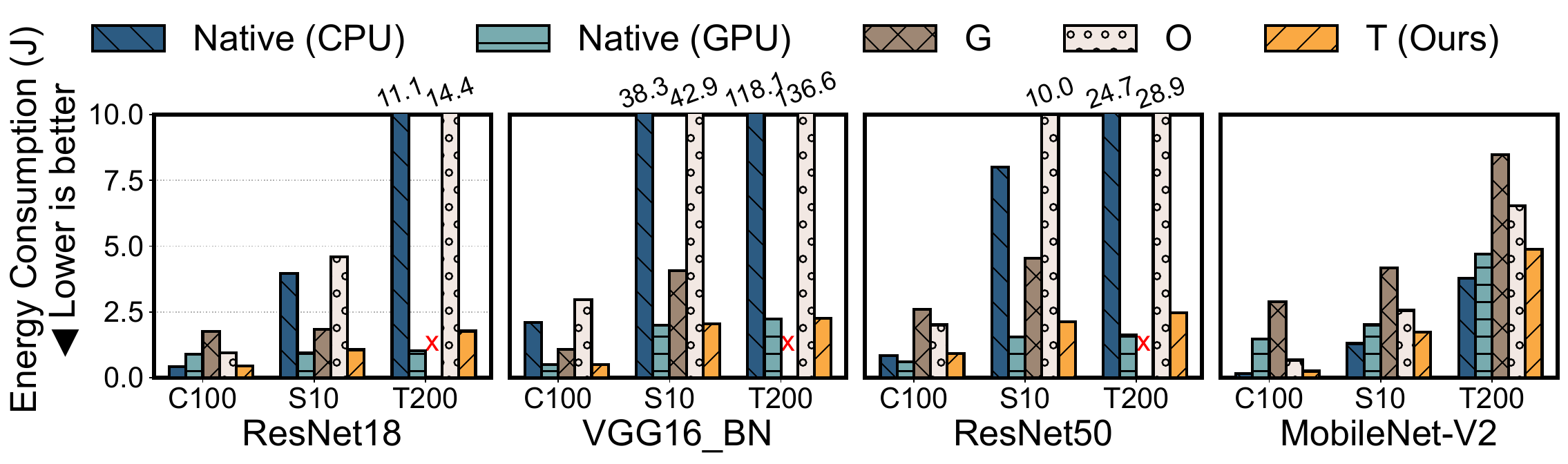}
    \end{minipage}
    \vspace{-1em}
    \begin{minipage}[t]{\linewidth}
    \centering
    \includegraphics[width=\linewidth]{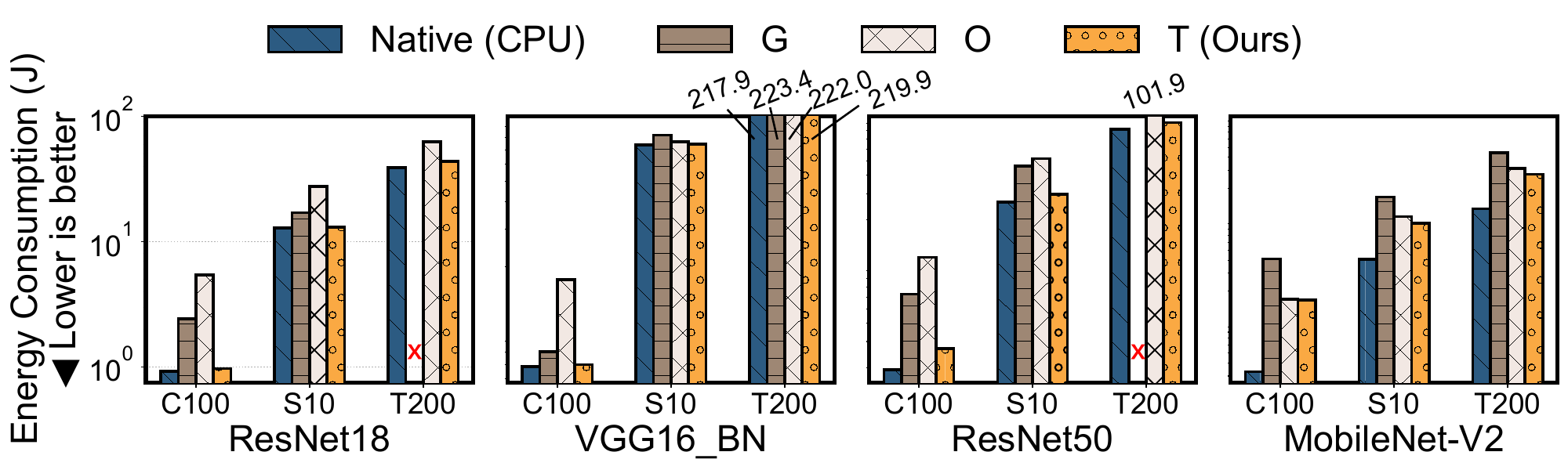}
    \end{minipage}
    \caption{Comparison of native and all-shield solutions with energy consumption. \textit{Left}: Hikey960. \textit{Right}: RPI3B+. "G", "O", and "T" represent GroupCover, Occlumency, and \name~(Ours), respectively. \ding{55} represents OOM.}
    \vspace{0em}
    \label{fig:eva_energy}
\end{figure}

We measure the energy consumption using a Deli DL333501C power meter~\cite{Deli} during the four model inference processes on two platforms.
The results are shown in Figure~\ref{fig:eva_energy}.
\name~achieves an average reduction in energy consumption of 58.66\% (up to 91.35\%) and 69.86\% (up to 98.35\%) compared to GroupCover and Occlumency, respectively, on the Hikey960. 
On the RPI3B+, it similarly reduces energy consumption by an average of 34.23\% (up to 60.05\%) compared to GroupCover and an average of 32.07\% (up to 82.09\%) compared to Occlumency. 
These improvements can be attributed to the reduced on-device inference time. 
Relative to native execution, \name~on the Hikey960 consumes, on average, 36.89\% less energy than native CPU execution, but 7.07\% more than native GPU execution. 
On the RPI3B+, it shows an average increase of 60.61\% in energy consumption compared to native CPU execution.

\subsection{Overhead}
\textbf{Hardware profiling time.}
We evaluate the overhead of \name's profiling time on two hardware platforms for computation modeling.
The total profiling times for all models on each device are detailed in Table~\ref{tab:eva_profiling}.
RPI3B+'s profiling time is significantly longer than Hikey960 due to the limited computation power.

\textbf{Tensor Evaluating time.}
Critical tensor evaluating is another major overhead in \name.
\name~requires a maximum of 20 epochs (average 10.9 epochs) compared to the standard evaluation method which uses 100 epochs. 
This results in an 89.1\% reduction in evaluation time while maintaining equivalent assessment accuracy, showing the efficacy of our selection algorithm (cf.~\ref{subsec:design_critical_tensor_selector}).

\textbf{Planning time.}
Since the impact of consecutive tensors can be neglected, our optimization solution time is less than 10 seconds, which is negligible.

Notably, the profiling is a one-shot offline process for the specific platform, while the evaluating and planning stages are one-shot offline processes for the specific victim model.

\begin{table}[t]
\centering
\caption{Total profiling times of two devices.}\label{tab:eva_profiling}
\vspace{-1em}
\resizebox{\linewidth}{!}{
\begin{tabular}{c|cccccc}
\Xhline{1pt}
\multirow{2}{*}{\textbf{Device}} & \multicolumn{6}{c}{\textbf{Profiling time}}                         \\ \cline{2-7} 
                        & \textbf{CPU}   & \textbf{GPU}  & \textbf{Transfer} & \textbf{Masking}  & \textbf{Obfuscating} & \textbf{Total}   \\ \hline
Hikey960~\cite{hikey960}               & 3min  & 3min & 6min     & 6min     & 5min        & 23min   \\ \hline
Raspberry Pi 3B+~\cite{rpi3b}        & 18min & /    & 42min & 1h 18min & 5min        & 1h 25min \\ \Xhline{1pt}
\end{tabular}
}
\vspace{-1em}
\end{table}

\section{Discussion}

\textbf{Supported models.}
In this paper, we focus on CNN models because they are well-deployed on devices.
For example, 
Xu et al.~\cite{xu2019first} comprehensively investigated 16,500 of the most popular Android apps on smartphones, and found that 87.7\% of the models used in deep learning apps are CNN models.
\name~can also be applied to other popular model architectures.
For example, deploying Large Language Models (LLMs) on mobile devices has achieved great advances~\cite{xue2024powerinfer, yin2024llm, xu2024fwdllm,cai2023federated}, and the transformer-based model also shows that different component (sub-layer) in trained Transformer models have different importance~\cite{wang2020rethinking}.
Expanding \name~to optimize for LLMs will be our future work.



\textbf{New TEE architectures.}
Despite ARM TrustZone has been widely used in the mobile and IoT industry, new TEE architectures (e.g., ARM CCA~\cite{ARM_CCA} and RISC-V KeyStone~\cite{lee2019keystone}) are still emerging.
Although such new TEEs may mitigate the performance overhead of shielding the entire model solutions, they do not harm the practicality of \name~because the computation speed of such new TEEs is still not comparable with the device's hardware accelerators (e.g., GPU).
We believe \name~can be a promising solution to bridge the gap between the new TEEs and the evolving accelerators.

\section{Related Work}
\begin{table}[t]
\centering
\caption{Comparison of \name~with related work}
\vspace{-1em}
\label{tab:related-work}
\resizebox{\linewidth}{!}{
\begin{threeparttable}[b]
\begin{tabular}{ccccc}
\Xhline{1pt}
\multicolumn{1}{c|}{\multirow{2}{*}{Works}} & \multicolumn{2}{c|}{Model Security}                & \multicolumn{1}{c|}{\multirow{2}{*}{Shielding}} & \multirow{2}{*}{Efficiency} \\ \cline{2-3}
\multicolumn{1}{c|}{}                       & \multicolumn{1}{c|}{MS} & \multicolumn{1}{c|}{MIA} & \multicolumn{1}{c|}{}                           &                              \\ 
\Xhline{1pt}
DarkneTZ~\cite{mo2020darknetz}&\halfcirc&\halfcirc&deep layers&High
\\
\rowcolor{gray!15}Serdab~\cite{elgamal2020serdab}& \halfcirc & \halfcirc & shallow layers & High 
\\
Magnitude~\cite{hou2021model} & \halfcirc& \halfcirc & large mag. weights& High\\
\rowcolor{gray!15}SOTER~\cite{shen2022soter} & \halfcirc& \halfcirc& Inter. layers & High\\
ShadowNet~\cite{sun2023shadownet} & \halfcirc & \halfcirc & non-linear layers& Medium \\
\rowcolor{gray!15}GroupCover~\cite{zhanggroupcover} & \fullcirc & \fullcirc  & non-linear layers & Medium\\
Occlumency$^\ddagger$~\cite{lee2019occlumency}&\fullcirc&\fullcirc&all layers& Low \\
\rowcolor{blue!15}\textbf{\name~(Ours)}& \fullcirc & \fullcirc & critical tensors & High\\
\Xhline{1pt}                                
\end{tabular}
\begin{tablenotes}
 \item[$\dagger$] \fullcirc~~~stands for fully protection. $^\ddagger$~It needs to encrypt weights offline.
\end{tablenotes}
\end{threeparttable}  
}
\vspace{-1em}
\end{table}
\quad \textbf{On-device secure inference.}
We show the related work in Table~\ref{tab:related-work}.
To fully protect model security, some previous work~\cite{lee2019occlumency, hou2021model, xie2024memory}
leverage TEE to shield the entire model to achieve black-box protection.
They propose advanced techniques to reduce secure memory usage but execute all layers in the TEE where lose the opportunity to leverage GPU for acceleration.
To speed up secure inference, existing research partitions the model and fixed part of the model to the TEE~\cite{elgamal2020serdab, mo2020darknetz, shen2022soter, hou2021model}.
Although they reduce inference latency, they cannot fully defend MS and MIA~\cite{zhang2023no}.
In order to enable the entire model to be accelerated by GPU, ShadowNet~\cite{sun2023shadownet} and GroupCover~\cite{zhanggroupcover} partition
the DNN by the layer types and shield non-linear layers using TEE. 
The offloaded linear layers are protected by lightweight obfuscation algorithms (e.g. matrix transformation) and all intermediate features are masked by OTP. 
The former utilizes a fixed obfuscation that has security issues while the latter utilizes random obfuscation achieving the black-box protection.
However, both of them have high overheads of obfuscation and masking.
Overall, none of the existing works simultaneously satisfy our requirements.



\textbf{Mitigating model attacks in training.}
Existing defenses have explored different training strategies for migrating the model attacks. 
For example, Beowulf~\cite{gong2024beowulf} mitigates MS attacks by reshaping the victim model's decision regions to more complex and
narrow.  
AMAO~\cite{jiang2023comprehensive} proposed to use adversarial training to shield victim models from MS.
SELENA~\cite{tang2022mitigating} mitigates MIA by self-distillation by a novel ensemble architecture in the training process.
TEESlice~\cite{zhang2023no} is a recent work that inserts small external model slices into the public model and only trains the slices to force private weights located in the slices.
Shielding these slices in the TEE can defend against MS and MIA.
However, the scenario of TensorShield (and previous TEE-based secure inference solutions) is different from the above work, i.e., performing security analysis after the victim model is finished trained in an arbitrary strategy.

\textbf{TEE in GPUs.}
Researchers have explored GPU TEEs to guarantee
data security on GPUs. 
Implementing trusted architectures directly inside GPUs could achieve isolation~\cite{volos2018graviton,hua2022guardnn},
but require customizing hardware (e.g., NVIDIA H100 GPU~\cite{H100GPU}) and are specific designed for cloud servers.
Recent works have also proposed several ARM-based GPU TEEs~\cite{deng2022strongbox,park2023safe, wang2024cage, wang2023secure}. 
However,
our solution employs GPUs in an ``out-of-the-box'' manner, which requires no change to the hardware or shipped firmware for mobile and IoT devices.

\section{Conclusion}
In this paper, we present \name, a new technique that selects critical tensors to fully defend against MS and MIA. 
\name\ achieves up to 25.35$\times$ (avg. 5.85$\times$) speedup in inference latency
without accuracy loss when compared to the state-of-the-art
schemes, and also reduces energy consumption by an average of 58.66\%.

\bibliographystyle{ACM-Reference-Format}
\bibliography{sample-base}

\end{document}